\documentclass{article}

\usepackage{arxiv}

\usepackage[utf8]{inputenc} 
\usepackage{amsmath}
\usepackage{bm}
\usepackage{amssymb}
\usepackage{amsfonts}
\usepackage{graphicx}
\usepackage{comment}
\usepackage{float}
\graphicspath{ {./Figures/} }
\usepackage{caption}
\usepackage{array}
\usepackage{geometry}
\usepackage{setspace}
\usepackage{fancyhdr}
\usepackage{subcaption}
\usepackage{booktabs}
\usepackage[svgnames,table]{xcolor}
\usepackage{adjustbox}
\usepackage{multirow}
\usepackage{hyperref}
\usepackage{soul}
\usepackage[backend=biber,style=numeric]{biblatex}
\title{Spatially-informed Image Harmonization Results in Improved Scanner Effect Removal and Prediction}

\author{
 Alec Reinhardt \\
  Department of Biostatistics\\
  The University of Texas MD Anderson Cancer Center\\
  Houston, TX 77030 \\
  \texttt{aereinh@unc.edu} \\
   \And
 Yajie Liu \\
  School of Public Health\\
  University of Texas Health Science Center at Houston\\
  Houston, TX 77030 \\
  \texttt{yajie.liu@uth.tmc.edu} \\
  \And
 Suprateek Kundu \\
  Department of Biostatistics\\
  The University of Texas MD Anderson Cancer Center\\
  Houston, TX 77030 \\
  \texttt{SKundu2@mdanderson.org} \\
}

\begin{document}
\maketitle
\begin{abstract}
976

\end{abstract}


\section{Introduction}

Classical neuroimaging studies are often limited by small or moderate sample sizes leading to inaccurate findings and compromised reproducibility \cite{marek2022reproducible}. The challenge of limited samples sizes are often compounded when the underlying true effect sizes are small  \cite{spisak2023multivariate}, as is often the case for neuroimaging data. 
Large multi-site studies such as ADNI, \cite{mueller2005alzheimer}
UK Biobank, \cite{sudlow2015uk} and the Human Connectome Project \cite{van2013wu} have arisen to tackle this problem by providing increased sample sizes designed to detect reproducible signals. Larger samples from these efforts enhance the power to detect significant associations and may improve the generalizability of results. However, these multi-site studies may inadvertently introduce site/scanner-specific heterogeneity (referred to as batch effects hereafter) in data acquisition and processing. Such batch effects may have an adverse impact by introducing unwanted variability that can mask significant associations as well as introduce spurious associations, and ultimately reduce the reliability of measurements. \cite{han2006reliability, kruggel2010impact, reig2009assessment, wonderlick2009reliability} 
Existing literature typically relies on data harmonization techniques such as ComBat \cite{fortin2018harmonization, fortin2016removing, rao2017predictive, marek2019identifying} to remove batch effects in the mean and variance terms, and subsequently use these harmonized images for downstream analysis. The goal of harmonization is to remove unwanted technical variation owing to inherent scanner differences, while preserving biological variation within the imaging signals. In doing so, a valid harmonization method can ensure comparability across multiple study sites and assist downstream statistical analyses by increasing power and eliminating variability unrelated to the biological analysis.

Unfortunately, the state-of-the-art harmonization methods essentially ignore the spatial information in the images when estimating the region/voxel-level scanner effects and other biological effects of interest in the harmonization model. It is known that ignoring such spatial correlations could result in inaccurate estimates in neuroimaging studies, particularly when the signal-to-noise ratio (SNR) is low. Existing literature illustrates that by accounting for spatial correlations for mean parameters in Bayesian generalized linear models, it is possible to obtain more accurate and reliable estimates \cite{mejia2020bayesian, spencer2022spatial}. Similarly, modeling of the local spatial dependence of the residual noise terms also increased statistical power and sensitivity \cite{bernal2013spatiotemporal,risk2016spatiotemporal}. The benefits of incorporating correlations in residual parameters when performing image harmonization was recently illustrated by \cite{chen2022mitigating} who incorporated site-specific correlated residual terms in the ComBat model. While providing notable improvements over existing ComBat approaches, this approach still did not address important gaps. In particular, the correlation structure was applied on the residual terms only but correlation between the mean parameters such as scanner effects was not accounted for. Moreover, the correlation structure did not explicitly account for spatial information in the images and certain modeling assumptions were imposed on the covariance parameters in order to make the approach computationally feasible that may be restrictive. Finally, the number of covariance parameters increased quadratically with the sample size that may result in overfitting issues when conducting harmonization for high-dimensional images involving tens of thousands of voxels.

We develop a novel Bayesian tensor based modeling approach for spatially aware voxel-level data harmonization for T1w-MRI scans derived from multi-site studies. The proposed harmonization approach is based on Bayesian tensor response regression models (BTRR) in statistical literature that is able to explicitly account for the spatial configuration of imaging voxels while simultaneously imposing dimension reduction to tackle possible overfitting.  By incorporating spatial information when modeling both the mean batch effects as well as residual terms, and subsequently removing scanner effects via a Combat style adjustment step, the proposed approach is able to garner sharp improvements compared to state-of-the-art methods. This is evident from our harmonization of T1w-MRI scans from over 2100 imaging scans from Alzheimer's Disease Neuroimaging Initiative study (ADNI-1) subjects, where the proposed approach yields harmonized images with improved (spatially distributed) scanner effect removal, accurate retention of biological signals, and greater reproducibility. A fundamental difference between proposed Tensor-ComBat (TC) approach and existing harmonization methods is the ability to perform inference on significant spatially distributed scanner effects as well as biological effects on the brain, under a fully Bayesian implementation. This feature simultaneously provides rich insights into the neurobiological mechanisms of Alzheimer's disease (AD) progression under an integrated framework that is not achievable under existing harmonization methods. It also provides a natural mechanism for uncertainty quantification, that is not straightforward based on empirical Bayes methods such as ComBat.

\section{Results}

\subsection{Method overview and benchmarking}

\begin{figure}[!h]
    \centering
    \includegraphics[width=\linewidth]{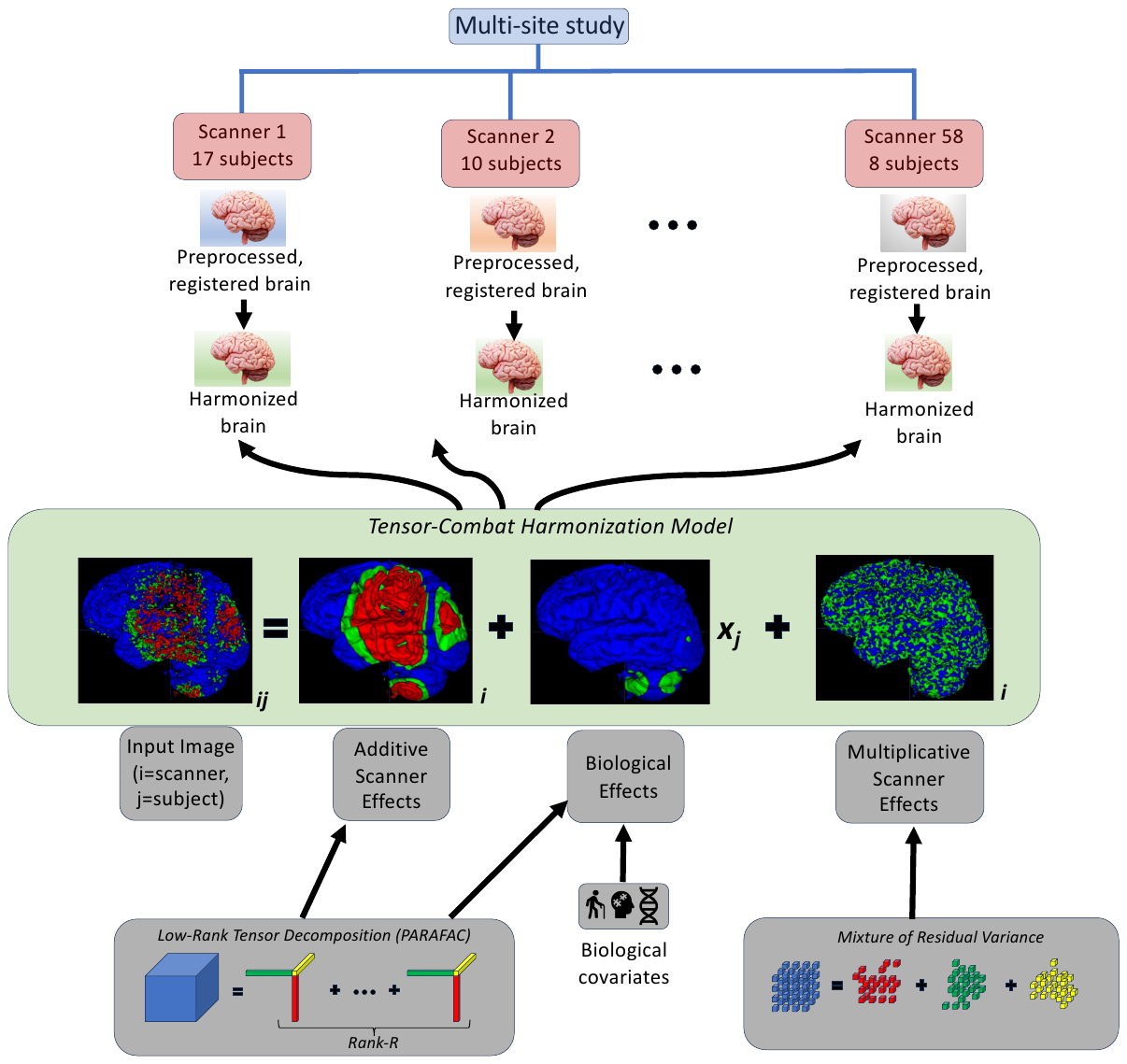}
    \caption{The pre-processing pipeline is illustrated in this Figure. }
    \label{fig:Fig1}
\end{figure}

The technical details for Tensor-ComBat (TC) are described fully in ``Methods" section, and a schematic with an overview of the approach is displayed in Figure \ref{fig:Fig1}. 
We present the harmonization results for cortical thickness (CT) maps derived from the T1w-MRI, from 2108 samples in the Alzheimer's disease neuroimaging initiative (ADNI-1) study as described in Section 2.2. We chose cortical thickness (CT) as our primary neuroimaging feature of interest, since they are known to be a highly sensitive biomarker for AD \cite{weston2016presymptomatic}. We validate our analysis by additionally performing the harmonization on the intensity maps observed directly from the T1w-MRI scans, as a second neuroimaging feature of interest.

 We validate the performance of Tensor-ComBat for harmonizing T1-weighted MRI neuroimaging data using various downstream analyses. These include quantifying scanner effect removal (G1) that is the primary metric for evaluating the success of any harmonization pipeline. We augment this classical goal by inferring spatially distributed brain regions with the highest scanner deviations that is a hallmark of the proposed TC method. Additionally, while scanner effect removal eliminates technical variability due to sites/scanner, the goal of harmonization approaches also include preserving the variability in the images due to biological signals (G2). This is evaluated via the prediction of  biological features from harmonized images, via a downstream analysis. We evaluate prediction accuracy corresponding to biological features included in the harmonization model, as well as cognitive scores that are not directly included in the harmonization step. In addition, we assess the generalizability of our findings by evaluating the reproducibility of spatially distributed brain regions that are associated with biological covariates (G3). 
 
We assessed downstream performance for multiple versions of Tensor-ComBat (TC) and alternative harmonization methods, focusing on scanner effect removal and prediction of biological signals using harmonized images. Competing approaches include the state-of-the-art ComBat (C), Adjusted-Residuals (AR), and the unharmonized images (U). Descriptions of these approaches along with technical details are provided in the Methods section. For all harmonization methods, we considered both cross-sectional (CS) case and longitudinal (L) harmonization models. We denote the cross sectional and longitudinal implementations for tensor combat as TC-CS and TC-L, and the corresponding methods for combat as C-CS and C-L respectively. 
Unlike the cross-sectional models that were fit by ignoring the within-subject dependence across longitudinal visits, longitudinal models controlled for within subject dependence via a subject-specific intercept term, along with linear temporal trend term, and time-covariate interactions. However, the advantages of the longitudinal analysis is somewhat mitigated by a larger number of model parameters compared to the cross-sectional analysis, which could result in challenges in model fitting. While 3-D analysis is desirable, it also contains a massive number of imaging voxels to be harmonized that pose computational challenges due to high-dimensionality. Therefore, it is appealing to harmonize the 2-D slices separately and then use these slices to reconstruct the 3-D image. We implement both variants of the model under Tensor-ComBat and denote them as 3-D harmonization (TC-CS-3D) and slice-by-slice (TC-CS-Sl) harmonization for the cross-sectional analysis and we denote the corresponding longitudinal versions as TC-L-3D (for 3-D) and TC-L-Sl (slice-by-slice) respectively. For ComBat, we harmonize the vectorize 3D images directly. 

\subsection{Materials and Data}

We analyzed the dataset consisting of longitudinal T1-weighted MRI images for $818$ subjects (with each individual having 1-3 three visits resulting in a total of 2108 images) across $58$ total study sites. Each site used a single MRI scanner and each subject visited a single study site for the course of the study. 
In addition, the ADNI-1 dataset included information on several demographic, biological, and cognitive features which are relevant for downstream analyses. Among these features, we included the following ones in the harmonization step: age, gender, diagnosis of Alzheimer's Disease, diagnosis of late mild cognitive impairment (LMCI), and presence of APOE-4 alleles (coded as 0 if APOE-4 is 0 and 1 otherwise).  The goal of including these features in the harmonization model was to evaluate whether the harmonized images was able to retain the biological signals corresponding to these covariates via downstream prediction, as per our aim (G2). In addition, we also evaluate the ability of the harmonized images to predict cognitive test scores in ADNI, although these scores were not explicitly included in the harmonization model due to the directionality of associations (test scores should be regressed on brain images, not the other way around). These scores included the Mini-Mental State Examination (MMSE), Rey Auditory Verbal Learning Test (RAVLT), and the Alzheimer's Disease Assessment Scale (ADAS) that are commonly used to evaluate cognitive impairment in AD. A summary of the ADNI-1 sample, stratified by disease group, is provided in Table \ref{tab:adni_summ}.

\begin{table}[]
    \centering
    \begin{tabular}{c|c|c|c|c}
    \toprule
         & {\bf AD} & {\bf LMCI} & {\bf Control} & {\bf Total}  \\
         {\bf Characteristic} & ($n=188$) & ($n=401$) & ($n=229$) & ($n=818$) \\
    \hline
         Number of sites & 52 & 57 & 55 & 58 \\
         Subjects per site [Median (Range)] & 4 (1-9) & 7 (2-19) & 4 (1-9) & 14 (2-37) \\
         Number of visits & 2.4 (0.9) & 2.6 (0.8) & 2.7 (0.7) & 2.6 (0.8) \\
         Age [Mean (SD)] & 75.3 (7.5) & 74.7 (7.4) & 75.9 (5.0) & 75.2 (6.8) \\ 
         Female & 89 (47\%) & 143 (36\%) & 110 (48\%) & 342 (42\%) \\
         $\geq 1$ APOE-4 allele & 124 (66\%) & 215 (54\%) & 61 (27\%) & 400 (49\%) \\
         MMSE [Mean (SD)] & 23.3 (2.0) & 27.0 (1.8) & 29.1 (1.0) & 26.7 (2.7) \\
         RAVLT [Mean (SD)] & 29.6 (10.0) & 38.6 (10.7) & 52.8 (10.0) & 40.5 (13.4) \\
         ADAS-13 [Mean (SD)] & 29.0 (7.6) & 18.7 (6.3) & 9.5 (4.2) & 18.4 (9.2) \\
    \bottomrule
    \end{tabular}
    \caption{Breakdown of the ADNI-1 dataset in terms of selected features for harmonization and downstream analysis.}
    \label{tab:adni_summ}
\end{table}

Before harmonization methods could be applied, voxel-level T1-w images were registered to a common template representative of the sample  (see ``Methods'' section). Subsequently, we applied a robust processing pipeline to segment and extract cortical thickness measures for all voxels of each image, given that cortical thickness is a relevant predictor of Alzheimer's Disease. The pre-processing workflow is described in detail in the Methods section and illustrated in Figure \ref{fig:Fig1}. Preprocessed images contained  $11,141,120$ voxels ($256\times 256\times 170$), which were downsampled to $27,000$ voxels ($30 \times 30 \times 30$). The downsampling step was consistent with classical approaches that perform harmonization at a macro level based in regions of interest (ROI) \cite{Johnson2007,Beer2020,Fortin2016}, and keeping in mind the computational burden incurred at such massive dimensions. However, the downsampled images are still at a much finer resolution compared to the ROI level harmonization that is usually performed in literature. We then applied a screening step to remove voxels which were were either outside of the brain mask or had zero variability in cortical thickness across subjects. This screening step removed impertinent voxels that should be excluded from the analysis. 



\subsection{Tensor-based Harmonization Provides Higher Scanner Effect Removal}

We evaluate the ability of different approaches for scanner effect removal using various metrics. First, it was of interest to quantify the amount of inter-scanner variation in CT measure before and after harmonization. Clearly, images with minimal scanner differences would exhibit greater similarity compared to images with inadequate scanner effect removal or unharmonized images. To this end, we examined pairwise similarity metrics (i.e., correlation and RMSE) between scanner-level mean cortical thickness (CT) images. Among the $\binom{58}{2}=1653$ pairs of scanners, we found that all examined harmonization methods led to significantly higher pairwise correlations and lower pairwise RMSEs compared to the unharmonized CT images. Additionally, we found that the slice-by-slice TC-CS method (TC-CS-Sl) applied to voxel-level data led to modest but significant improvements compared with cross-sectional ComBat for both pairwise correlation ($0.2\%$ increase, $p<2e-16$, BH-corrected t-test) and RMSE ($7\%$ decrease, $p=3.4 \times 10^{-10}$, BH-corrected t-test). {\bf ROI-Level Results:} We repeated this analysis using ROI-level harmonized outcomes, which were derived from the voxel-level harmonized images for the TC methods, and obtained with competing methods by harmonizing the ROI-level CT directly. We find that the TC-CS-Sl method leads to significantly higher pairwise correlation in ROI-level CT compared with C-CS ($0.8\%$ increase, $p=2.82\times 10^{-17}$, BH-corrected t-test). TC-CS-Sl method also shows significant improvements with respect to pairwise correlation compared with the next-closest method, AR-CS ($0.5\%$ increase, $p=1.72\times 10^{-6}$, BH-corrected t-test), and all other competing approaches. Similarly, we see that TC-CS leads to significantly lower pairwise RMSE under ROI-level harmonization compared with C-CS ($14.4\%$ decrease, $p=0.008$, BH-corrected t-test), AR-CS ($15.2\%$ decrease, $p=0.005$, BH-corrected t-test), and all other competing approaches. By contrast, TC-CS-3D only shows significant improvements for pairwise RMSE compared with C-CS ($12.8\%$ decrease, $p=0.02$, BH-corrected t-test) and AR-CS ($13.6\%$ decrease, $p=0.01$, BH-corrected t-test), but not for pairwise correlations. We observe that all methods except for C-L, TC-L-3D, and TC-L-CS) lead to significant improvements in pairwise correlation and RMSE over the unharmonized case. Based on these results, we find that the TC method, which harmonizes at the voxel-level and subsequently converts to ROI-level, leads to improved scanner effect removal over competing methods which harmonize directly at the ROI-level.

On the other hand, neither the 3D TC-CS method nor the longitudinal variations of Tensor-ComBat showed significant improvements in pairwise scanner similarity over cross-sectional ComBat. The benefits for scanner removal using the TC-CS-Sl over the 3D version can be attributed to the fact that the slice-by-slice harmonization models are more efficient to fit, and thus can accommodate higher choices of rank which are still computationally feasible. In contrast to the cross-sectional methods, corresponding longitudinal methods (for TC, ComBat, and adjusted-residuals) resulted in weaker improvements in pairwise similarity metrics over the unharmonized case. Greater variation in pairwise metrics were also observed for longitudinal methods. 
The relatively poor performance of the longitudinal Tensor-Combat approach compared to the cross-sectional counterpart may be potentially due to the presence of voxel-specific and subject-specific random intercept terms that introduces a large number of parameters, which may not be precisely estimated from our longitudinal ADNI data with only up to three observed visits per subject. A greater number of visits per subject may be desirable for producing improved harmonization results under the longitudinal version of the model.

\begin{figure}[!h]
    \centering
    \includegraphics[width=\linewidth]{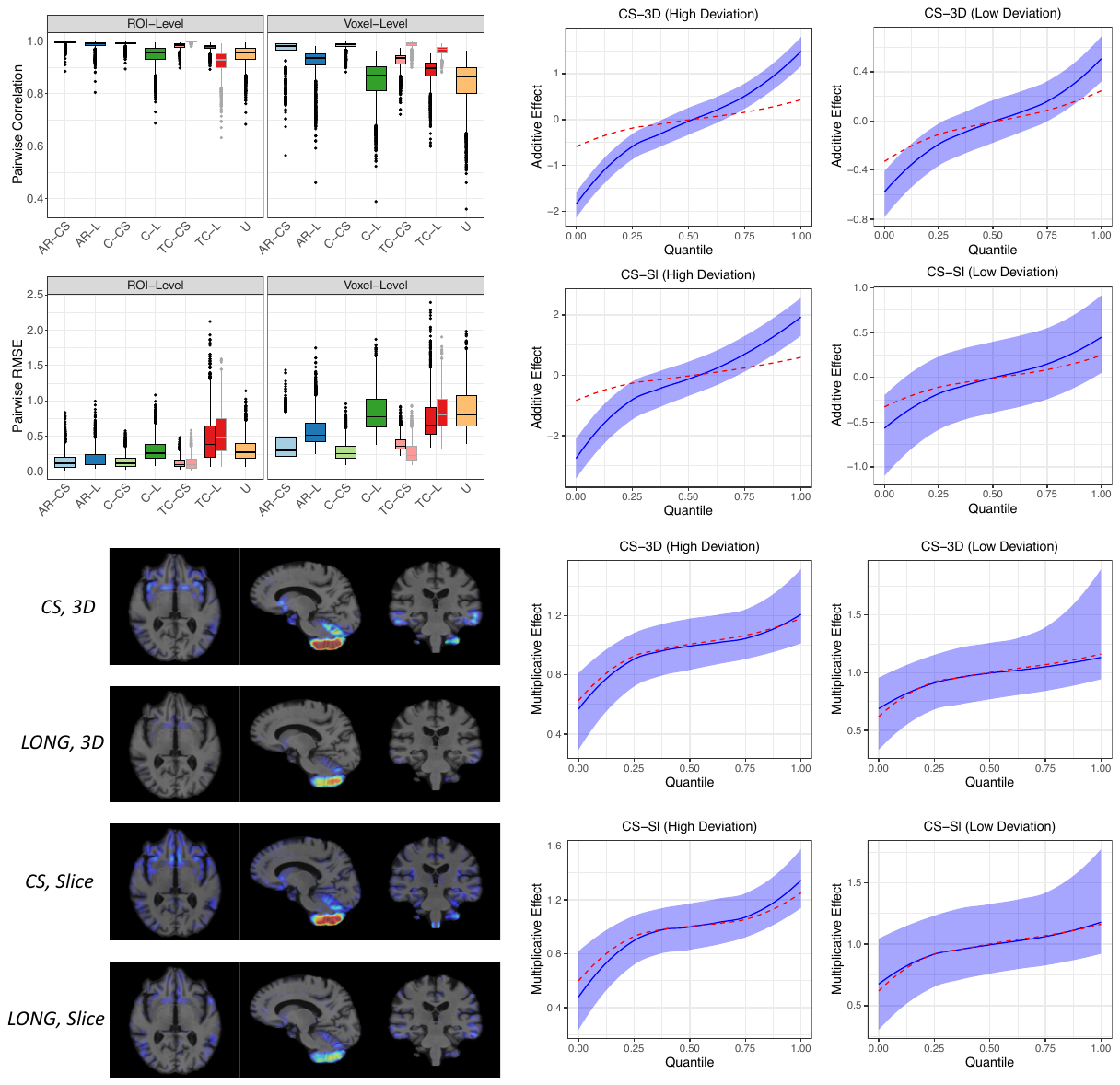}
    \caption{For distribution plots (left half), we define a region of ``high deviation'' as all voxels such that $>10\%$ of scanner pairs }
    \label{fig:Fig2}
\end{figure}

\underline{Spatially-localized scanner effects}: One  novel feature of the Tensor-ComBat approach, compared with state-of-the-art alternatives like ComBat, is the capacity to quantify uncertainty about spatially distributed voxel-level deviations across scanners. This natural uncertainty quantification arises from the fully Bayesian implementation of the Tensor-ComBat harmonization model, which allows for the computation of multiplicity-corrected credible regions in the scanner effect estimates across voxels  (see ``Methods'' for details). By contrast, the state-of-the-art ComBat method is typically fit via empirical Bayes that requires ad hoc steps such as bootstrapping to generate uncertainty estimates for scanner effects. Using the Tensor-ComBat method, we performed pair-wise scanner comparisons at each voxel to identify voxels where each pair of scanners showed significant differences. We then report a brain map showing the proportion of pairwise scanner comparisons that show significant additive differences averaged across voxels (Figure \ref{fig:Fig2}). We see that the most prominent inter-scanner differences appear around the cerebellum, with 36\% of scanner pairs reporting significant difference when averaged across voxels. This is notable, given that the cerebellum reports the highest variation in cortical thickness across individuals in the ADNI-1 dataset that may potentially result from a significant proportion of voxels lying in the cerebellum (11\%) compared to other ROIs that contain at most 5\% of voxels (see Table \ref{tab:roi_volume} in Supplementary Materials). Outside of the cerebellum, the inferior temporal gyrus and fusiform gyrus were found to have the highest proportion of significant deviations across scanners, using the TC-CS-Sl method to control for biological and spatial dependence. In particular, 10.7\% of scanner pairs showed significant differences in the left inferior temporal gyrus and 11.5\% in the right inferior temporal gyrus, averaged across voxels in the respective ROIs. For the fusiform gyrus, 9.3\% of scanner pairs displayed a difference in the left hemisphere and 7.6\% in the right hemisphere, averaged across voxels in each ROI.



\underline{Distributions of scanner effects}: In addition to spatial maps, we compared the distributions of additive and multiplicative scanner effects between the Tensor-ComBat model and ComBat. We assessed these distributions separately for regions with high estimated scanner deviations and for the entire cortical mask. The resulting distributions, plotted as quantile curves in Figure \ref{fig:Fig2} indicate that additive scanner effect estimates varied significantly between Tensor-ComBat and ComBat for localized regions with high scanner deviations using a Kolmogorov Smirnoff test ($p=0.027$). However, the distributions across the entire brain mask for the additive scanner effects did not vary significantly between Tensor-ComBat and ComBat ($p=0.167$), nor did the distributions for multiplicative effects either within the high-deviation region ($p=0.203$) or the entire mask ($p=0.289$). 
Moreover, we find that estimated spatial differences in additive scanner effects are relatively consistent across the four variations of Tensor-ComBat examined (cross-sectional and longitudinal, 3D and slice-by-slice), although cross-sectional methods yielded slightly higher degree of scanner deviations around the cerebellum region. Further, the higher scanner differences are primarily driven by higher values of additive scanner effects, while the multiplicative effects are relatively stable between the regions with high and low scanner variations.
These results indicate that Tensor-ComBat allows for more pronounced additive scanner corrections in spatially-localized signals than ComBat, while retaining similar scanner effect corrections across the entire image. These results provide rich insights into the mechanics of scanner differences under Tensor-ComBat, particularly when images contain sparse, clustered regions of technical variation. 

\underline{Residual distributions by scanner}:
As another means to assess presence of scanner effects while considering potential downstream biological analysis, we examined the distributions of residuals after regressing each harmonized image set on biological covariates (i.e. age, AD, LMCI, gender, and APOE-4 alleles) using BTRR models. The resulting residual distributions are plotted by scanner in Figure \ref{fig:Fig3}. From these residuals, we found that both ComBat and TC-CS-Sl images had less than $0.5\%$ of voxels with significant differences in mean cortical thickness across scanners, while for the unharmonized images, $92\%$ of voxels showed evidence of significant differences based on ANOVA. Similarly, for differences in variance across scanners, we found that both ComBat and Tensor-ComBat had less than $4\%$ of voxels with significant differences, while the unharmonized images had $52\%$ of voxels with significant differences based on Bartlett's test. However, we find that Tensor-ComBat demonstrates substantial benefits over ComBat in terms of the goodness-of-fit and proportion of residual outliers. In particular, the mean $R^2$ value across voxels and scanners for TC-CS-Sl was $0.75$, which was significantly higher than the mean $R^2$ for ComBat of $0.63$ (student t-test, $p=0.00015$). 
The average percentage of outliers across all scanners was also significantly lower for Tensor-ComBat than for ComBat ($p=5.2 \times 10^{-8}$ under a binomial proportion test).  Our analysis illustrates that while both Combat and Tensor-ComBat results in adequate removal of scanner effects from the residual distributions across voxels, the ComBat approach produces large outliers for a non-trivial proportion of voxels in the harmonized image that can prove problematic for downstream analysis (see Sections 2.4-2.6). This finding is corroborated by a better goodness of fit under the proposed Tensor-ComBat approach compared to Combat.

\underline{Potential factors affecting scanner differences}

\begin{figure}[!h]
    \centering
    \includegraphics[width=\linewidth]{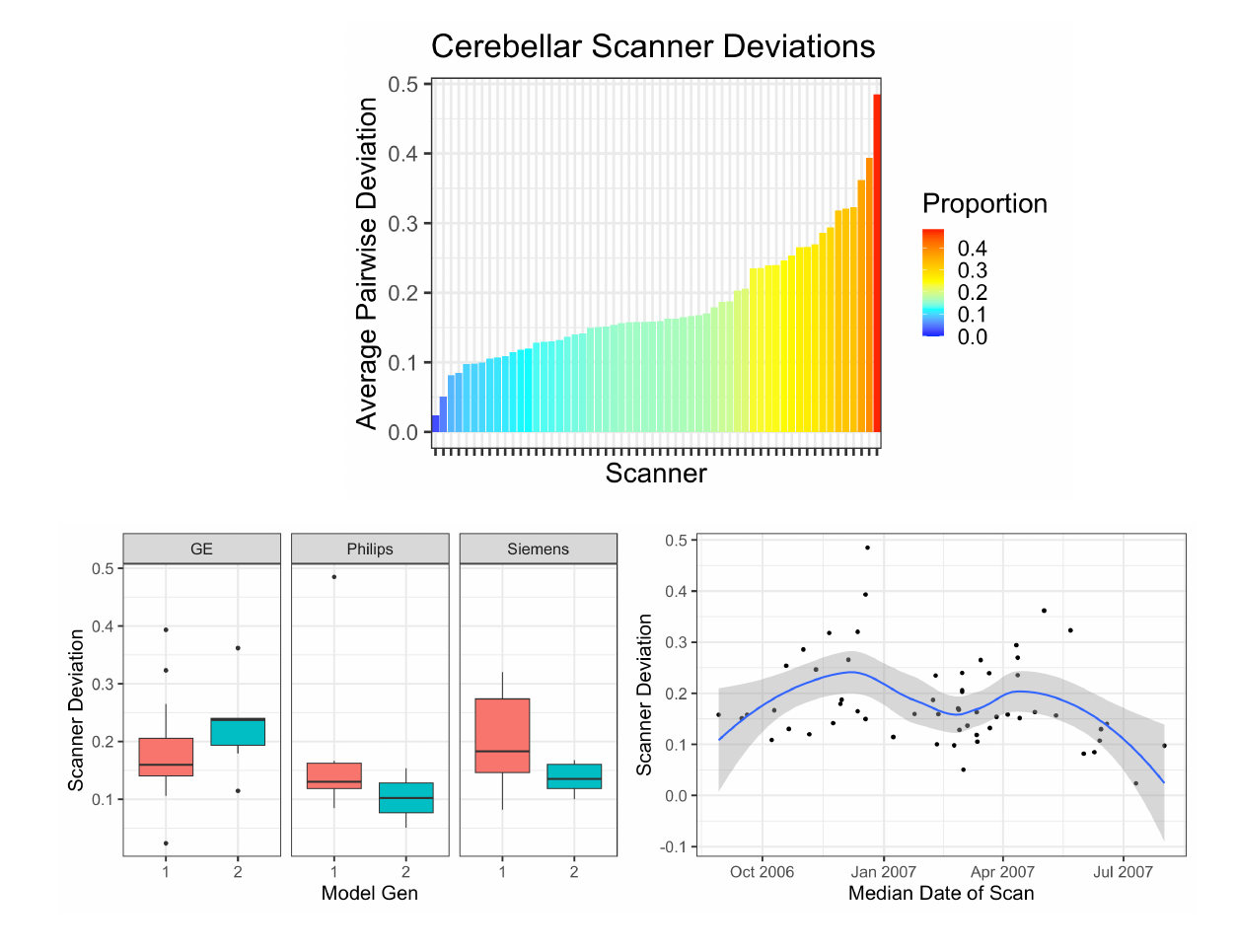}
    \caption{Generation 1 Models: Genesis Signa, Signa Excite (GE), Gyroscan NT, Gyroscan Intera, Intera (Philips), Sonata, SonataVision, Symphony (Siemens). Generation 2 Models: Signa HDx (GE), Intera Achieva, Achieva (Philips), SymphonyTim, Avanto, NUMARIS/4 (Siemens)}
    \label{fig:Fig_ScanDevbyType}
\end{figure}

\begin{figure}[!h]
    \centering
    \includegraphics[width=\linewidth]{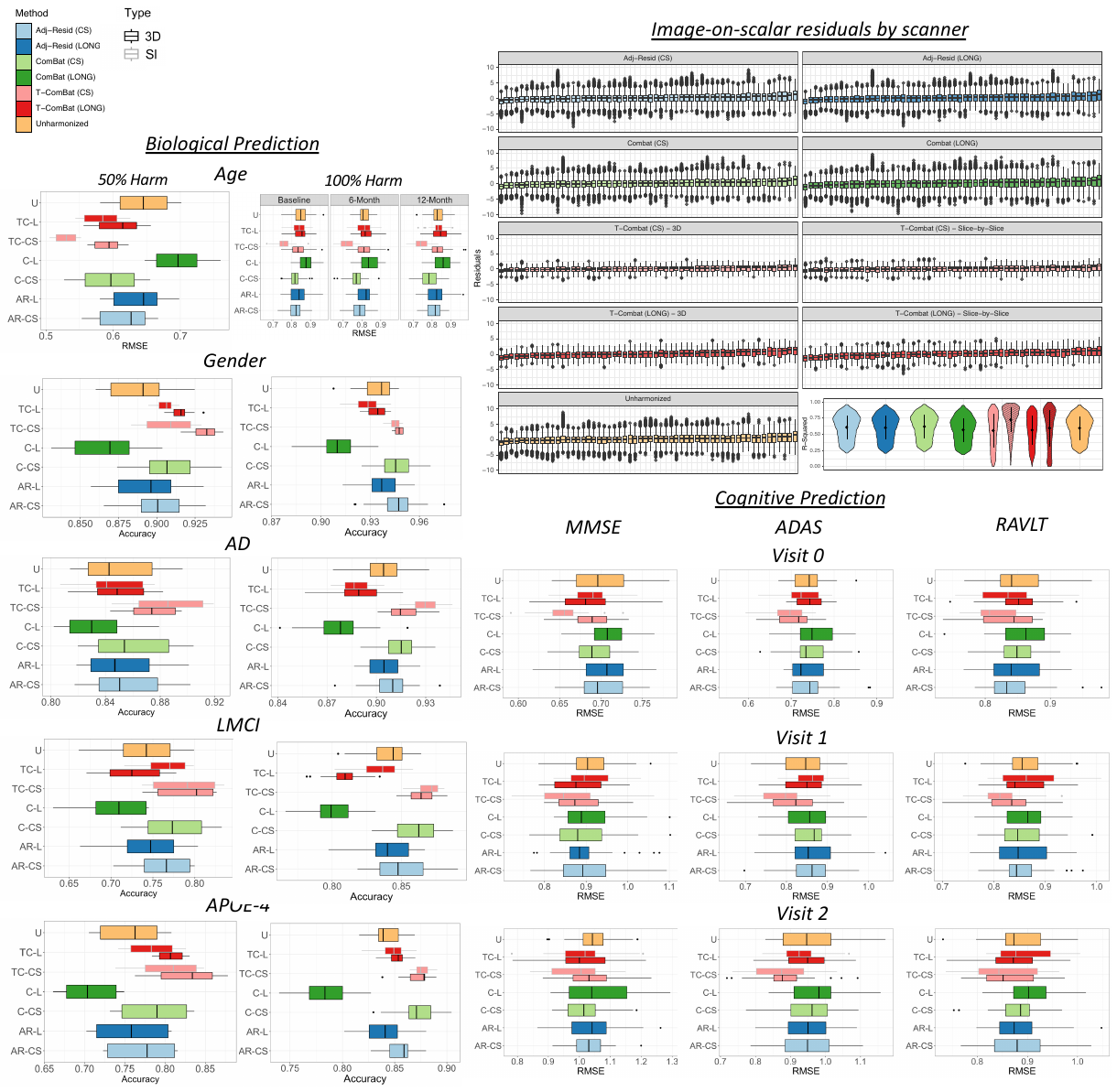}
    \caption{Biological (left) and cognitive scores (bottom right) prediction using unharmonized and harmonized images, fit to age, gender, cognitive syndrome diagnosis, APOE4 and cognitive scores. And image-on-scalar residuals by scanner.}
    \label{fig:Fig3}
\end{figure}

We explored how the estimated spatially-varying scanner differences varied across scanner type. In particular, we first defined a measure to characterize how much each scanner contributed to scanner differences in the cerebellar CT, using the proportion of significant pairwise differences in additive scanner effects, averaged across voxels in the cerebellum (top left panel of Figure \ref{fig:Fig_ScanDevbyType}). We characterize how these deviations vary across scanner manufacturer, model generation, number of samples within that scanner, and median scan date in Figure \ref{fig:Fig_ScanDevbyType}. In particular, we find that for Siemens scanners, there is a significant decrease ($p=0.012$, t-test) in scanner deviation from generation 1 models (Sonata/Vision, Symphony) to generation 2 models (Avanto, NUMARIS/4), though we do not find significant differences across model generation for the GE or Philips manufacturers. We also qualitatively observe that scanner deviations tend to increase when the number of samples associated with that scanner increases (need to comment on this in terms of what it means for model fitting), and the scanner deviations tend to be lower for scanners which were utilized later in the study.

\subsection{Tensor-based Harmonization Amplifies Biological Signals}

After harmonization, successful scanner effect removal should ideally be accompanied by improved ability to pick up on biological signals of interest in downstream analyses. To quantify these improvements, we performed a series of prediction tasks, including 5-fold cross-validation on all harmonized samples, in addition to a sensitivity analysis where a randomized 50\% subset of samples were harmonized and used for prediction. We explored both the prediction of biological variables included in the harmonization models, namely age, gender, clinical diagnosis, and APOE-4, in addition to cognitive features not directly accounted for during harmonization (MMSE, RAVLT, and ADAS). The results for prediction are displayed in Figure \ref{fig:Fig3}. We find that the slice-by-slice cross-sectional Tensor Combat (TC-CS-Sl) method leads to significant improvements over cross-sectional ComBat (C-CS) for out-of-sample prediction of age diagnosis ($p_{\text{adj}}=8.5e-19$), AD diagnosis ($p_{\text{adj}}=6.8e-6$) and LMCI diagnosis ($p=0.022$, $p_{\text{adj}}=0.2$). For the LMCI diagnosis, the significant improvements under TC-CS-Sl does not survive multiplicity adjustments. The strongest improvements were observed under TC-CS-Sl versus the unharmonized case for all 5 examined biological outcomes ($p_{\text{adj}}=1.09e-50$ for age, $p_{\text{adj}}=2.1e-4$ for gender, $p_{\text{adj}}=3.15e-8$ for AD, $p_{\text{adj}}=7.69e-9$ for LMCI, and $p_{\text{adj}}=9.43e-12$ for APOE-4).  
The full set of p-values comparing all pairs of methods for biological prediction is reported in the Supplementary Materials (Tables \ref{tab:agePred_s}-\ref{tab:apoePred_s})

Comparing the performance of four variations of the Tensor-ComBat model (TC-CS-Sl, TC-CS-3D, TC-L-Sl, and TC-L-3D), we note that TC-CS-Sl tends to lead to optimal out-of-sample prediction for the examined biological outcomes. In particular, we observe that TC-CS-Sl produces significantly lower out-of-sample RMSE for age prediction compared to all other TC models ($p_{\text{adj}}=8.02e-51$ compared with next-best case, TC-L-Sl) and AD prediction ($p_{\text{adj}}=2.62e-6$ compared with next-best case, TC-CS-3D). For gender prediction, we observe that TC-CS-3D outperforms TC-CS-Sl by a slight ($<1\%$) margin, though this difference was found to be significant ($p_{\text{adj}}=0.013$). For LMCI prediction, we find no significant differences between performance of TC-CS-Sl and TC-CS-3D ($p_{\text{adj}}=0.059$), and likewise for APOE-4 prediction ($p_{\text{adj}}=0.18$). 

Examining the longitudinal variants of TC, we see poor overall predictive performance compared to alternative approaches like TC-CS and cross-sectional Combat. In fact, compared with the unharmonized case, we only observe that TC-L (i.e., TC-L-3D) leads to significant improvements in prediction for APOE-4 allele status ($p_{\text{adj}}=0.008$), and none of the other examined biological covariates. Nonetheless, it is of note that TC-L outperforms longitudinal ComBat for prediction of all examined biological outcomes. Given the limited longitudinal information in this ADNI-1 imaging dataset, it may be of interest in future work to use richer longitudinal datasets and simulations to compare performance of longitudinal harmonization models.

{\bf Prediction with 50\% Subset}: We repeated harmonization and the 5-fold cross-validation biological prediction scheme on a randomly-selected subset of 50\% of the cohort. The goal of this analysis was to determine if the predictive advantages under the TC approaches were still retained for lower training sample sizes. 
We find that TC-CS-Sl leads to significant predictive gains in age prediction over C-CS ($p=0.002$ unadjusted, $p_{\text{adj}}=0.04$) and over all other competing approaches. We also that TC-CS-Sl leads to significant gains over the unharmonized case for AD prediction ($p=0.009$ unadjusted) and LMCI prediction ($p=0.02$ unadjusted), though these do not retain significance after multiplicity corrections (adjusted p-values of $p_{\text{adj}}=0.29$ and $p_{\text{adj}}=0.63$, respectively). Although the TC-CS-Sl approach results in considerable improvements in prediction accuracy compared to the other methods (including C-CS) for the other prediction scenarios, these improvements are not significant under multiplicity adjustments. This may indicate the need for different choices of tensor rank and other hyper-parameters to achieve a more optimal predictive performance for studies with smaller sample sizes under the proposed TC approach. However, the loss of significant improvements is a general trend for smaller sample sizes that is consistent throughout even when comparing different pairs of methods implemented. Although significant predictive improvements may not be achieved for all scenarios under TC methods for reduced sample sizes, the optimal performance under the TC approach for the full dataset provides one with confidence for implementing these approaches for smaller datasets.



All significance estimates for cross-validation results, including the full data and 50\% cohort with unadjusted (t-test), and adjusted (BH-corrected) p-values are included in Supplementary Tables \ref{tab:agePred_s}-\ref{tab:apoePred_s}.

\subsection{Tensor-based Harmonization of Improves Cognitive Prediction}

 We repeated the prediction analysis outlined above to evaluate the out-of-sample prediction accuracy for cognitive variables that were not directly included in the harmonization model. Predicting these cognitive features is of strong interest in literature, since AD is known to accelerate cognitive impairment (REF!!!). However unlike  in the previous section where the biological features of interest could be directly included in the harmonization model followed by downstream prediction, it is not possible to incorporate the cognitive features in the harmonization analysis due to violation of the direction of causality. In other words, one may not regress the brain images on cognitive variables, since cognitive abilities are expected to be influenced by brain structure, not the other way around. In Figure \ref{fig:Fig3}, we plot the out-of-sample RMSEs from 5-fold cross validation for predicting MMSE, RAVLT, and ADAS scores based on harmonized images. While the differences between unharmonized and harmonized method out-of-sample predictive performance are not as pronounced as in the case of prediction of biological variables, we do note the presence of some significant differences when predicting cognitive outcomes at each longitudinal visit separately.
 In particular, we find that TC-CS-Sl leads to significant gains in MMSE prediction over C-CS when predicting with images at baseline ($p=0.00013$ unadjusted, $p=0.004$ with BH-adjusted), but not when using images at either follow-up time-point. For ADAS prediction, we find significant predictive gains of TC-CS-Sl over C-CS for all timepoints (unadjusted p-values: $p=0.0009$, $p=0.0007$, and $p=0.001$ , adjusted p-values: $p=0.03$, $p=0.02$, and $p=0.04$, for baseline, m-06, and m-12, respectively). However, we do not find significant differences in out-of-sample RMSEs between TC-CS-Sl and C-CS for RAVLT prediction at any time point. We note that the largest relative difference in mean RMSEs between TC-CS-Sl and C-CS was for ADAS year-1 follow-up prediction ($8.7\%$ drop in RMSE for TC-CS-Sl vs. C-CS). For the TC-CS-3D, we find significant improvements in mean out-of-sample RMSE over C-CS for MMSE prediction at baseline and ADAS prediction at 6-month and 12-month follow-ups ({\it p-values table to be inserted}). We still see improvements comparable results under the TC-CS-3D method compared to C-CS for the other cognitive prediction cases, although the differences may not be significant. All the other harmonization approaches, including longitudinal TC method, did not show significant improvements compared to the unharmonized case. This illustrates the difficulty in predicting biological signals not directly included in the harmonization model.





\subsection{Tensor-based image harmonization results in greater reproducibility of biological signals}
\begin{figure}[!h]
    \centering
    \includegraphics[width=\linewidth]{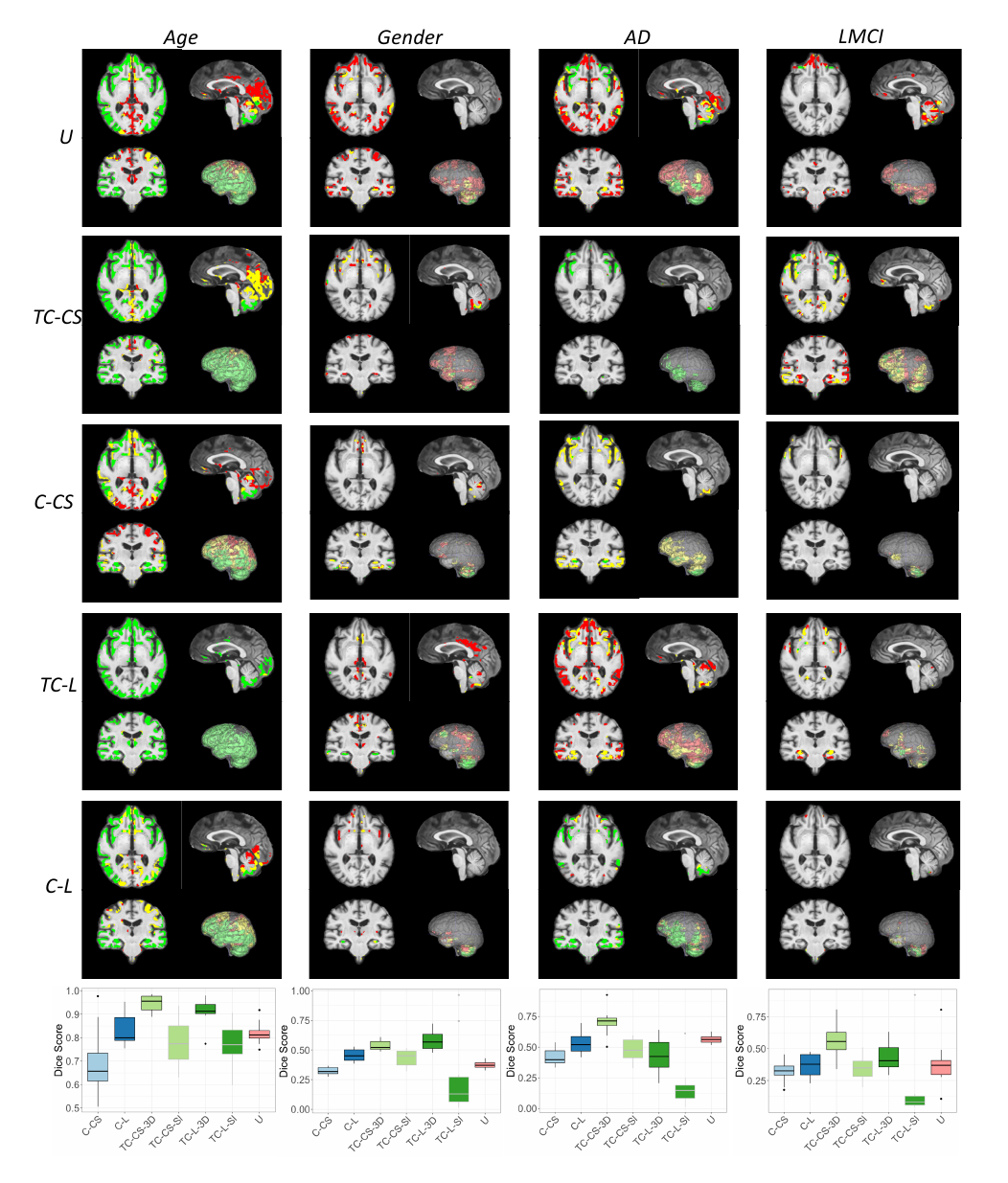}
    \caption{The overlapping significance maps and Dice scores for age, gender, cognitive syndrome diagnosis (AD or MCI).}
    \label{fig:Fig4}
\end{figure}

 A major concern in neuroimaging studies is the ability to obtain reproducible results (REF!!!).
Therefore, we evaluate the degree to which each harmonization method is able to produce reproducible results. In particular, we split the full dataset into $2$ randomly-selected subsets (replicates) of (roughly) half the sample size, 
that were approximately matched in terms of  the frequencies of scanners as well as other covariates included in harmonization pipeline. We then use a BTRR model to regress each set of CT images (unharmonized and harmonized) on each biological feature of interest, in order to obtain spatially distributed voxel-level estimates for the significant effects of biological features. Significance estimates are obtained using a joint credible intervals approach. The corresponding 3D significance maps for each replicate are obtained, and maps from each pair of replicates are compared against the full dataset model fit to get Dice scores, measuring overlap of significant regions. The overlapping significance maps and Dice scores are shown in Figure \ref{fig:Fig4} for five different biological features. We observe that the Tensor-Combat method has substantially higher mean Dice score for the significance maps pertaining to the effect of age, AD status, LMCI status,  on cortical thickness, compared with alternative harmonized image sets including ComBat. In particular, we find that the 3D cross-sectional Tensor-ComBat method tends to generally produce the highest average Dice scores (0.96 for age, 0.46 for gender, 0.71 for AD, 0.79 for LMCI), while the 3D longitudinal Tensor-ComBat produced comparable Dice scores (0.81 for age, 0.59 for gender, 0.41 for AD, 0.44 for LMCI) and cross-sectional TC with slice-by-slice analysis produced slightly lower scores ($0.78$ for age, $0.42$ for gender, $0.73$ for AD, $0.37$ for LMCI). In contrast,  ComBat  produced lower Dice scores compared to the all TC variants, except for gender where longitudinal ComBat has a higher Dice score compared to 3D cross-sectional TC (but lower reproducibility compared to 3D longitudinal TC). Moreover, longitudinal ComBat has greater Dice scores compared to CS-ComBat (0.81 for age, 0.57 for gender, 0.52 for AD, 0.39 for LMCI). For all methods, the brain maps for APOE-4 had sparsely populated significant effects that led to a low reproducibility throughout.

Although all variants of Tensor-ComBat had higher or comparable Dice coefficients compared to ComBat for age, these improvements were not statistically significant under our scheme involving 5 replicates. Moreover, while the longitudinal 3D Tensor-ComBat has higher Dice scores for compared to ComBat variants for gender effect estimation, the difference is not statistically significant. Moreover, longitudinal ComBat has higher reproducibility compared to other TC variants. On the other hand, for both AD and LMCI effect size estimation, cross-sectional TC variants produced significantly higher Dice coefficients compared to ComBat variants.

\newpage

\section{Discussion}

{\underline{Cognitive prediction:}}However, it is worth noting that the relative magnitudes of these differences was not as prominent as the case of prediction biological factors that were also explicitly included within the harmonization model. These results suggest that harmonization can lead to slight amplification of signals which are not accounted for directly during harmonization, but which are implicitly addressed due to underlying correlations between biological factors included in the harmonization pipeline. 

{\underline{Reproducibility:}}
Coupled with the fact that TC-CS-Sl has the best predictive performance followed by 3D TC-CS, the high reproducibility of these approaches points to a robust and reliable harmonization approach. Noting that the TC-CS-Sl method also excels in terms of scanner effect removal, this harmonization technique may be considered as the best choice for harmonization for this data.

\section{Methods}

\subsection{Overview of harmonization for neuroimaging data}

Our general setup involves a multi-site neuroimaging study that collects structural images (e.g., T1-weighted MRI) on $n$ subjects using $K$ different imaging scanners ($1 < K < n$). Studies may be cross-sectional or longitudinal with varying numbers of visits $T_i$ for each subject. For our purposes, we assume that a given subject is assigned to one scanner for the span of the study that is also the case for our ADNI-1 dataset. All images across subjects and timepoints are registered to a common template such that each preprocessed image contains $p$ voxels distributed over identical spatial locations. 
Our goal is to harmonize spatially distributed neuroimaging features such as cortical thickness derived from observed T1w-MRI image intensities. 
Further, let us denote the image at the $t$-visit as $\mathcal{Y}_{it}$, denote $\{x_1,\ldots,x_q \}$ as the set of biological covariates to be incorporated in the harmonization model, and let $\text{vec}(\cdot)$ denote the vectorization operation for a given tensor object. We denote the scanner label associated with the $i$-th subject as $S_i$ that could indicate any of the labels in $[1,K]$.

The classical harmonization approach that controls for additive and multiplicative scanner effects, along with biological covariates, can be encapsulated by the following statistical model:
\begin{align}
    \mathcal{Y}_{it} = \mathcal{M} + \sum_{s=1}^q \Theta_s x_{is} + \Gamma_{S_i} + \mathcal{E}_{it}, \label{eq:basemodel}
\end{align}

\noindent where $\mathcal{M}$ is the population-level intercept, $\Theta_s$ is the effect pertaining to biological covariate $\mathbf{x}_s (s=1,\ldots,S)$, and $\Gamma_{S_i}$ is the additive scanner effect corresponding to scanner $S_i$. Scanners are also assumed to have multiplicative effects which are heterogeneous across voxels. These multiplicative effects are included in the tensor-valued residual term $\mathcal{E}_{it}$, whose $v$-th element is given as $\epsilon_{v,it} \sim N(0,  \sigma_{\epsilon,v S_i}^2)$, where $\sigma_{\epsilon,vS_i}^2$ is the site- and voxel-specific residual variance term. Often, it is assumed that $\sigma_{\epsilon, v S_i} = \delta_{v,S_i} \sigma_{\epsilon,v}$, where $\sigma_{\epsilon,v}$ is the common variance terms across sites/scanners for voxel $v$, and $\delta_{v,S_i}$ is the scanner-specific residual variance scaling term (i.e. multiplicative scanner effect) for voxel $v$.

\subsection{Brief review of existing harmonization approaches}

We describe some standard approaches for removal of scanners effects in literature, all of which are based on linear regression. First, we will introduce some notations. For the sake of simplicity, we ignore the notations for longitudinal visits for now, but include these notations later when describing the longitudinal harmonization model. Denote the imaging feature $v$ ($v$ may index different ROIs or voxels) corresponding to the $i$th subject and $j$th site/scanner as $y_{ijv}, v=1,\ldots,V,j=1,\ldots,K,i=1,\ldots,n_j$, assuming that there are $n_j$ subjects for the $j$th site ($j=1,\ldots,K$). Further, denote the $n_j\times 1$ vector of the $v$-th imaging feature across all samples at the $j$-th site as ${\bf y}_{jv}$, where $v=1,\ldots,V$.  We will assume that measurements are collected for $q$ biological covariates of interest that are common across all subjects and sites, and denoted as ${\bf x}_{i}$ corresponding to the $i$th subject. These variables are understood to contain clinical, behavioral and demographic features. The corresponding $n_j\times p$ covariate matrix across all samples for the $j$th site is denoted as $X_{j}$. Further, denote the effect of the $j$-th site on the $v$-th imaging feature as $\gamma_{jv}$ that is estimated by pooling information across all samples. 

{\noindent \underline{Unadjusted Residuals:}} This is one of the simplest harmonization approaches that adjusts for site effects, but does not adjust for any other biological covariates of interest. The method first fits the following linear regression model corresponding to the $v$th imaging feature: $
{\bf y}_{jv} = \alpha_v{\bf 1}_{n_j} + \gamma_{jv}{\bf 1}_{n_j} + {\bm\epsilon}_{jv}, \mbox{ } j=1,\ldots,K,$
and subsequently computes the harmonized residuals as ${\bf y}^{UA}_{jv} = {\bf y}_{jv}  - \hat{\gamma}_{jv}{\bf 1}_{n_j}$, where the parameter estimates $\gamma_{jv}$ are computed via ordinary least squares (OLS) separately for each imaging feature. This approach is known as {\it unadjusted residuals} and may suffer from inadequate harmonization due to the lack of biological covariates and due to independent estimation across voxels without pooling information across the image. 

{\noindent \underline{Adjusted Residuals:}}
This approach removes site effects after adjusting for biological covariates using the following multivariable regression model 
\begin{eqnarray}
{\bf y}_{jv} = \alpha_v{\bf 1}_{n_j} + X_j {\bm\beta}_v + \gamma_{jv}{\bf 1}_{n_j} + {\bm\epsilon}_{jv}, \mbox{ } j=1,\ldots, K, \label{eq:adj-res}
\end{eqnarray}
where ${\bf 1}_{n}$ denotes a vector of ones of length $n$, $\alpha_v$ is the voxel/ROI-specific intercept term, ${\bm\beta}_v$ is the $q\times 1$ vector of effects corresponding to the $q$ biological predictors, $\gamma_{jv}$ is the voxel/ROI specific and site-specific effect accounting for heterogeneity across scanner sites for a given feature $v$, and $\epsilon's$ are independently distributed Gaussian errors with residual variance $\sigma^2_\epsilon$. 
The estimated parameters ($\hat{\alpha_v}, \hat{{\bm\beta}_v}, \hat{\gamma_{jv}}$) are subsequently used to compute the harmonized adjusted residuals as ${\bf y}^A_{jv} = {\bf y}_{jv}  - \hat{\gamma}_{jv}{\bf 1}_{n_j} - X_j \hat{\beta}_v$. Given that this approach is able to account for variability due to biological covariates of interest while simultaneously adjusting for site/scanner effects, it is able to cater to confounding caused due to potential interactions between these two factors. The model is fit using the OLS approach applied separately to each voxel, which does not allow one to leverage spatial information within the image.

{\noindent \underline{ComBat Approach:}} The ComBat approach is currently state-of-the-art for harmonization. It is essentially an empirical Bayes approach that uses a variant of model (\ref{eq:adj-res}) by incorporating a feature- and site-specific scaling factor and fitting the model jointly across all voxels. The model is defined as follows -
\begin{eqnarray}
{\bf y}_{jv} = \alpha_v{\bf 1}_{n_j} + X_j {\bm\beta}_v + \gamma_{jv}{\bf 1}_{n_j} + \delta_{jv}{\bm\epsilon}_{jv}, \mbox{ } {\bm\epsilon}_{jv}\sim N(0,\sigma^2_{\epsilon,v}I_{n_j}), v=1,\ldots,V, j=1,\ldots,K , \label{eq:combat}
\end{eqnarray}
where the residual errors follow a Gaussian distribution, and suitable prior distributions are specified on the model parameters as: ${\bm\beta}_v\sim N(0,\sigma^2_b), \gamma_{jv}\sim N(\gamma_j,\sigma^2_\gamma), \delta^2_{jv}\sim IG(\lambda_j,\theta_j)$, under a Bayesian set-up. Here $\gamma_{j} $ denotes the common prior mean for $\{\gamma_{j1},\ldots, \gamma_{jv} \}$ towards which all site-specific parameters effects are shrunk, $IG$ denotes an inverse Gamma distribution, and $N(\mu,\sigma^2)$ denotes a Gaussian distribution with mean parameter $\mu$ and variance $\sigma^2$. ComBat uses empirical Bayes to estimate the parameters $(\hat{\gamma},\hat{\delta},\hat{{\bm\beta}},\hat{\alpha}_v)$ jointly across all features/voxels $v=1,\ldots,V,$ and sites $j=1,\ldots,K,$ via a multi-step procedure that relies on closed form posterior distributions. The corresponding harmonized residuals are obtained as: $y^{Combat}_{ijv} =  \frac{1}{\hat{\delta}_{jv}}\big(y_{ijv}  - \hat{\gamma}_{jv} - \hat{\alpha}_v - {\bf x}^T_{ijv} \hat{{\bm\beta}}_v  \big) + \hat{\alpha}_v + {\bf x}^T_{ijv} \hat{{\bm\beta}}_v $ based on the point estimates under the empirical Bayesian approach.  However, while this approach provides improvements over the OLS estimators for adjusted and unadjusted residuals, it still treats the ROIs/voxels as interchangeable and hence suffers from not being able to incorporate the spatial information in the image. Similar to the OLS estimators, it is not straightforward to report uncertainty and perform inference for ComBat. 




\subsection{Tensor-Combat}
We developed the harmonization methodology for the Tensor-ComBat approach based on a Bayesian tensor response regression model (BTRR) that is applicable to spatially distributed voxel-wise imaging features for both 3-D and 2-D brain images. When applying this proposed method to two-dimensional (2-D) imaging slices, we first harmonize various 2-D slices derived from the 3-D brain image, and subsequently re-stack the harmonized 2-D slices to reconstruct the harmonized 3-D image. Below, we will describe the methodology for 3-D images, but the method and notations can be generalized to 2-D images in a straightforward manner.

Consider the pre-processed 3-D T1-MRI brain images for the $j$th scanner and $i$th individual as a tensor object that is denoted as $\mathcal{Y}_{ij}\in \Re^{p_1\times p_2\times p_3}$ for the $i$-th sample ($i=1,\ldots,n_j,$), with $p_1,p_2,p_3,$ voxels along the three dimensions. The pre-processing steps are described elsewhere in the Methods section. For a given voxel $v$, the corresponding measurement is denoted as $y_{ij}(v)$ or alternatively $y_{ijv}$. We propose the following BTRR model, which views the image as a tensor-valued object, and models the imaging features jointly across all voxels as: 
 \begin{eqnarray}
\mathcal{Y}_{ij} = \mathcal{M} +\sum_{s=1}^q \Theta_s x_{is} + \Gamma_j + \mathcal{E}_{ij}, j=1,\ldots,K, i=1,\ldots, n_j,\mbox{ } \quad \label{eq:BTRR}
\end{eqnarray}
 where $\mathcal{M}$ is the population-level intercept, $\Theta_s$ is the effect pertaining to biological covariate $x_s (s=1,\ldots,S)$, and $\Gamma_{S_i}$ is the additive scanner effect corresponding to scanner $S_i$.  Scanners are also assumed to have site-specific variances that are heterogeneous across voxels. These effects are accounted for via the tensor-valued residual term $\mathcal{E}_{it}$, whose $v$-th element is given as $\epsilon_{v,it} \sim N(0,  \sigma_{\epsilon,{jv}}^2)$, where $\sigma_{\epsilon,{jv}}^2$ is the common residual variance across all scanners and subjects for a given voxel $v$, and $\delta_{v,S_i}$ is the scanner-specific residual variance scaling term (i.e. multiplicative scanner effect) for voxel $v$. Here, all the coefficients in the above model are tensor-valued and belong to the space $\Re^{p_{1}\times p_{2}\times p_{3}}$. In particular, the tensor-valued coefficients follow a low-dimensional PARAFAC decomposition that expresses them as an outer product of tensor margins as:
$\Theta_{s} = \sum_{r=1}^R \theta_{1\bullet,rs} \circ  \theta_{2\bullet,rs} \circ  \theta_{3\bullet,rs}$, $\Gamma_j = \sum_{r=1}^R \gamma_{1\bullet,rj} \circ \gamma_{2\bullet,rj} \circ \gamma_{3\bullet,rj}$, $\mathcal{M} = \sum_{r=1}^R \mu_{1\bullet,r} \circ  \mu_{2\bullet,r} \circ  \mu_{3\bullet,r}$, where $\circ$ denotes an outer product and $R$ denotes the rank of the tensor.

{\noindent \underline{Interpretation of tensor specification:}} The intercept and the coefficient terms in model  (\ref{eq:BTRR}) are expressed as outer products of tensor margins summed over $R$ channels, where $R$ is considered the rank of the tensor decomposition. For example, $\theta_{d\bullet,rs} (p_d\times 1)$ denotes the tensor margin for mode $d$ ($d=1,2,3$) corresponding to the $r$th channel and $s$th covariates. These tensor margins are combined in a multiplicative manner via an outer product and summed over $R$ channels to recover the regression coefficient tensor $\Theta_{s} $. This low rank representation is known as the PARAFAC decomposition, which is able to massively reduce the dimension of the parameter space from $p_1\times p_2\times p_3$ distinct parameters to only $R(p_1+p_2+p_3)$ for each tensor coefficient in the model. Such a tensor decomposition leads to a massive reduction in the number of parameters compared to a voxel-wise analysis or even compared to ComBat methods, which avoids overfitting and makes the proposed approach computationally scalable to tens of thousand of voxels. Moreover, while the same value of the tensor rank $R$ is used for all coefficient matrices in model (\ref{eq:BTRR}) in our implementation, this assumption can be relaxed to induce coefficient-specific ranks. 

We note that model \ref{eq:BTRR} may be extended to model longitudinal images by incorporating a subject-specific intercept and time-varying covariates using the following model:
 \begin{eqnarray}
\mathcal{Y}_{ijt} = \mathcal{M} + B_i + \sum_{s=1}^q \Theta_s x_{is} + \Gamma_j + \mathcal{E}_{ijt}, \mbox{ } j=1,\ldots,K, i=1,\ldots, n_j, t=1,\ldots,T, \label{eq:BTRR_long}
\end{eqnarray}

where $t=1,\ldots, T,$ indexes longitudinal visits, $B_i$ denotes the tensor-valued random intercept term that accounts for the within-subject dependence across visits, tensor coefficients $\{\mathcal{M},\Theta_1,\ldots, \Theta_q\}$ are defined similarly as in model ({\ref{eq:BTRR}}), $\mathcal{E}_{ijt}$ denotes residual error terms corresponding to visit $t$, and similarly for site effects $\Gamma_j,j=1,\ldots,J$.




{\noindent \underline{Prior Specification:}} We adapt the recent developments in longitudinal BTRR models \cite{kundu2023bayesian} for our analysis involving multi-site studies with a large number of samples. The prior specifications corresponding to the parameters in (\ref{eq:BTRR}) are as follows:
\begin{eqnarray}
&& \gamma_{d\bullet,rj} \sim \mathcal{N}\left(\mathbf{0},\tau^{\gamma} W^{\gamma}_{d,rj}\right),  
\mbox{ }
\theta_{d\bullet,rs} \sim \mathcal{N}\left(\mathbf{0},\tau^{\theta}_{s} W^{\theta}_{d,r}\right), \mbox{ } 
 \mu_{d\bullet,r} \sim \mathcal{N}\left(\mathbf{0},\tau^{\mu} W^{\mu}_{d,r}\right),  \nonumber \\
&&  \sigma^2_{\epsilon,{jv}}\sim \sum_{h=1}^H \pi_{\sigma,h}\delta_{\zeta_h}, \mbox{ } \zeta_h\sim IG(a_\epsilon, b_\epsilon), \mbox{ } \tau^\gamma \sim \text{Ga}(a_{\tau},b_{\tau}),  \tau_s^\theta \sim \text{Ga}(a_{\tau},b_{\tau}), \tau^\mu \sim \text{Ga}(a_{\tau},b_{\tau}), \mbox{  } \label{eq:BTRRprior}
\end{eqnarray}
\noindent where $\delta_{x}$ denotes a point mass at $x$, $IG(\cdot,\cdot)$ and $Ga(\cdot,\cdot)$ refer to the inverse Gamma and Gamma distributions respectively, and the site and voxel-specific residual variance terms $\sigma^2_{\epsilon,iv}$ follow a mixture distribution that specifies clusters of voxels sharing identical (but unknown) residual variance terms resulting in dimension reduction. The cluster memberships as well as values of the atoms within the cluster are learnt in an unsupervised manner, under an inverse Gamma prior on the cluster atoms. 
The number of mixture components $H$ is pre-specified and subject to hyperparameter tuning. Further, we specify a suitable structured prior covariance $W_{d,r} (p_d\times p_d)$ for the $d$th tensor margin designed to induce explicit spatial prior correlations between the regression coefficients corresponding to neighboring voxels. This is especially useful to infer clusters of spatially distributed voxels that correspond to significantly associations between the image and covariates, and is expected to be more flexible compared to an independence assumption on the tensor margins assumed in literature. In particular, we specify $W^{\gamma}_{d,rj} = \mbox{diag}(\sqrt{w^{\gamma}_{d,rj,1}},\ldots, \sqrt{w^{\gamma}_{d,rj,p_d}}) \times \Lambda \times  \mbox{diag}(\sqrt{w^{\gamma}_{d,rj,1}},\ldots, \sqrt{w^{\gamma}_{d,rj,p_d}})$, where the matrix $\Lambda$ imposes spatial correlations with a AR-1 structure having a lengthscale parameter $\alpha^{\gamma}_{d,r}$. For further simplification and ease in posterior computation, the $p_d$ diagonal variance terms are assumed to be equal, i.e. $w^{\gamma}_{d,rj,1}=\ldots=w^{\gamma}_{d,rj,p_d}=w^{\gamma}_{d,rj}$, with the following hierarchical priors imposed:
$w^{\gamma}_{d,rj}\sim \text{Exp}(\lambda_{d,r}^{\gamma}/2), \ \lambda_{d,r}^{\gamma}\sim \text{Gamma}(a_{\lambda},b_{\lambda})$. We request the readers to refer to \cite{kundu2023bayesian} for more details on the model, and insights into the properties of tensors.

\vskip 10pt

\vskip 10pt

{\noindent \underline{Data Harmonization Procedure}:} 
Using a similar adjustment strategy as for ComBat method, the harmonized image responses under Tensor-ComBat are given as
\begin{eqnarray}
y^{TC}_{ijv} = \frac{y_{ijv} - \widehat{\mathcal{M}}(v) - \sum_{s=1}^q \widehat{\Theta}_s(v) x_{is} - \widehat{\Gamma}_j(v)}{\widehat{\delta}_{\epsilon,jv}} + \widehat{\mathcal{M}}(v) + \sum_{s=1}^q \widehat{\Theta}_s(v) x_{is}, \quad \label{eq:tensorcombat}
\end{eqnarray}
where the superscript $TC$ refers to the proposed Tensor-ComBat approach and $(\widehat{\mathcal{M}}(v),\widehat{\Theta}(v),\widehat{\Gamma}_i(v),\widehat{\delta}_{\epsilon,jv})$ denote the estimated parameters corresponding to the $v$th imaging feature, obtained from the posterior distribution. The residual variance scaling term $\widehat{\delta}_{\epsilon,jv}$ is obtained post-hoc from the sampled scanner-specific residual variances. In particular, it is defined as the ratio of scanner-specific variances and the weighted mean variance over all scanners, i.e. $\widehat{\delta}_{\epsilon,jv}=\frac{\widehat{\sigma}_{\epsilon,jv}}{\widehat{\sigma}_{\epsilon,v}}$, where $\widehat{\sigma}_{\epsilon,v} = \frac{\sum_{j=1}^K n_j \widehat{\sigma}_{\epsilon,jv}}{\sum_{j=1}^K n_j}$. These estimates correspond to the posterior means and are obtained by fitting the model jointly across all voxels under a Markov chain Monte Carlo (MCMC) computation scheme. Although the harmonized residuals follow a similar approach as when harmonizing the data under the ComBat method, there are considerable difference between these methods in terms of the modeling details and prior specifications. 

\subsection{Posterior Computation}
To implement the Tensor ComBat method, we use an Markov Chain Monte Carlo (MCMC) approach, which samples from the conditional posteriors of coefficient tensor margins and other model parameters in order to obtain point and uncertainty estimates of the harmonization coefficients. The MCMC steps for most parameters such as the tensor margins, scale parameters, and noise variance mixture components can be sampled efficiently using fully Gibbs updates with closed form posteriors, while the lengthscale parameters ($\alpha$) accounting for the intra-margin correlations are updated using Metropolis-Hastings. In particular, we use the proposal density $\alpha_{d,r,s_x+1}|\alpha_{d,r,s_x} \sim \text{log-Normal}(\alpha_{d,r,s_x},\sigma_{\alpha}^2)$, where $s_x$ indexes MCMC iteration and $\sigma_{\alpha}^2$ is a fixed variance term. After specifying hyperparameters, the choice of coefficient rank, and number of mixture components for $\sigma_{\epsilon,iv}^2$, and initializing model parameters (different choices for initialization -- priors, low-rank margins for Combat/adjusted residuals coefficients, e.g.), conditional posteriors are iteratively sampled from using the following procedure, looping over model terms (i.e. $\mathcal{M}$, $\Theta_s$, and $\Gamma_i$), rank channels ($r: 1\leq r \leq R$), and tensor dimensions ($d=1,2$ for each image slice):

\vspace{10pt}

\noindent {\it \emph{Step}} 1. Let $\widehat{\mathcal{Y}}_{ij,r}^{\mu}=\mathcal{Y}_{ij}-\sum_{s=1}^q \widehat{\Theta}_s x_{js}-\widehat{\Gamma}_i-\sum_{\substack{r'=1 \\ r'\neq r}}^R \mu_{1\bullet,r'} \circ \mu_{2\bullet,r'} \circ \mu_{3\bullet,r'}$ be the rank-specific residual for the population intercept $\mathcal{M}$. The $k$th element for margin $\mu_{d\bullet,r}$ for $k \in [2,p_d-1]$ follows the conditional posterior

\begin{align*}
    & \pi(\mu_{d\bullet,r,k}|-) = \mathcal{N}
    \left(\nu_{drk}^{\mu} \left[n_{drk}^{\mu} + \frac{e^{-\alpha_{dr}^{\mu}}(\mu_{d\bullet,r,k-1}+\mu_{d\bullet,r,k+1})}{\tau^{\mu} w_{dr}^{\mu} (1-e^{-2\alpha_{dr}^{\mu}})} \right], \ \nu_{drk}^{\mu} \right)
\end{align*}

\noindent where $\nu_{drk}^{\mu} = \frac{\tau^{\mu} w_{dr}^{\mu} (1-e^{-2\alpha_{dr}^{\mu}})}{m_{drk}^{\mu}\tau^{\mu} w_{dr}^{\mu} (1-e^{-2\alpha_{dr}^{\mu}}) + (1+e^{-2\alpha_{dr}^{\mu}})}$, $m_{drk}^{\mu} = \sum_{i,j} \|\frac{L_{dr}^{\mu}}{\sigma_{\epsilon,i(\ldots,k,\ldots)}}\|^2$, $n_{drk}^{\mu}=\sum_{i,j} \bigg\langle \frac{\widehat{\mathcal{Y}}_{ij,r}^{\mu}(\ldots,k,\ldots)}{\sigma_{\epsilon,i(\ldots,k,\ldots)}^2}, L_{dr}^{\mu} \bigg\rangle$, and $L_{dr}^{\mu}=\mu_{1\bullet,r}\circ \ldots \circ \mu_{d-1\bullet,r} \circ \mu_{d+1\bullet,r}\circ\ldots \circ \mu_{D\bullet,r}$, with $(\ldots,k,\ldots)$ representing the mode-$d$ matricization of a $D$-way tensor obtained by fixing the $d$th dimension at element $k$.

For $k \in \{1,p_d\}$, the conditional posterior is similar to the above, but with the $(1+e^{-2\alpha_{dr}^{\mu}})$ term in the expression for $\nu_{drk}^{\mu}$ replaced with $1$ and the $(\mu_{d\bullet,r,k-1}+\mu_{d\bullet,r,k+1})$ term in the posterior mean replaced with $\mu_{d\bullet,r,k+1}$ for $k=1$ and $\mu_{d\bullet,r,k-1}$ for $k=p_d$.

\vspace{10pt}

\noindent {\it \emph{Step}} 2. The diagonal entries of the margin covariance, $w_{dr}^{\mu}$ follows the conditional posterior

\begin{align*}
    & \pi(w_{dr}^{\mu}|-) = \text{GIG}\left(1-\frac{p_d}{2}, c_{dr}^{\mu}, \lambda_{dr}^{\mu} \right)
\end{align*}

\noindent where \text{GIG} denotes a Generalized Inverse Gaussian distribution, and \newline $c_{dr}^{\mu}=\frac{1}{\tau^{\mu}(1-e^{-2\alpha_{dr}^{\mu}})}\left[\mu_{d\bullet,r,1}^2+\mu_{d\bullet,r,p_d}^2 + (1+e^{-2\alpha_{dr}^{\mu}}) \sum_{k'=2}^{p_d-1} \mu_{d\bullet,r,k'}^2 - 2e^{\alpha_{dr}^{\mu}} \sum_{k''=1}^{p_d-1} \mu_{d\bullet,r,k''} \mu_{d\bullet,r,k''+1} \right]$.

\vspace{10pt}

\noindent {\it \emph{Step}} 3. The rate parameter $\lambda_{dr}^{\mu}$ follows the conditional posterior

\begin{align*}
    & \pi(\lambda_{dr}^{\mu}|-) = \text{Gamma}\left(a_{\lambda}+p_d, b_{\lambda}+\frac{p_d w_{dr}^{\mu}}{2} \right)
\end{align*}

\vspace{10pt}

\noindent {\it \emph{Step}} 4. The scale parameter $\tau^{\mu}$ follows the conditional posterior

\begin{align*}
    & \pi(\tau^{\mu}|-) = \text{GIG}\left(a_{\tau}-\frac{R(p_1+\ldots+p_D)}{2}, \sum_r \sum_d \mu_{d\bullet,r}^T W_{dr}^{-1} \mu_{d\bullet,r}, 2b_{\tau}\right)
\end{align*}

\noindent {\it \emph{Step}} 5. For the conditional posterior of lengthscale parameter $\alpha_{dr}^{\mu}$, we have the following:

\begin{align*}
    & \pi(\alpha_{dr}^{\mu}|-) \propto \left(\alpha_{dr}^{\mu}\right)^{a_{\alpha}-1} \left(1-e^{-2\alpha_{dr}^{\mu}}\right)^{-\frac{p_d-1}{2}} \exp\left[-\frac{1}{2} \mu_{d\bullet,r}^T W_{dr}^{-1} \mu_{d\bullet,r} + 2b_{\alpha} \alpha_{dr}^{\mu} \right]
\end{align*}

Using the above fact along with the Log-normal proposal density for $\alpha_{dr,s_x+1}^{\mu}|\alpha_{dr,s_x}^{\mu}$, we update $\alpha_{dr}^{\mu}$ using a Metropolis-Hastings step.

\noindent {\it \emph{Steps}} 6-15. The conditional posteriors for tensor margins, margin covariances (diagonal entries, scale parameters, and lengthscale parameters), and the rate parameters corresponding to the $\Theta_s$ and $\Gamma_i$ coefficients are updated in a similar manner as Steps 1-5, but with adjusted rank-specific residuals and, in the case of $\Theta_s$, incorporating vector covariates in the tensor margin conditional posterior as follows (with $s$ indexing covariate):

\begin{align*}
    & m_{drk,s}^{\theta} = \sum_{i,j} x_{ij,s}^2 \bigg \|\frac{L_{dr,s}^{\theta}}{\sigma_{\epsilon,i(\ldots,k,\ldots)}}\bigg \|^2 \\
    & n_{drk,s}^{\theta} = \sum_{i,j} x_{ij,s} \bigg \langle \frac{\widehat{\mathcal{Y}}_{ij,r,s}^{\theta}(\ldots,k,\ldots)}{\sigma_{\epsilon,i(\ldots,k,\ldots)}^2}, L_{dr,s}^{\theta} \bigg \rangle
\end{align*}

\noindent {\it \emph{Step}} 16. After updating all model coefficient parameters, we update the noise variance $\sigma_{\epsilon,i}^2$ via its mixture components $s_{ih}^2$ for site $i$ and component $h: 1\leq h \leq H$, whose prior probabilities are given by $\boldsymbol{\pi}_i = (\pi_{i,1},\ldots,\pi_{i,H})$. After obtaining the residual term $\mathcal{R}_{ij}=\mathcal{Y}_{ij}-\widehat{\mathcal{M}}-\sum_{s=1}^Q \widehat{\Theta_s} x_{ij,s} - \widehat{\Gamma}_i$, we first sample from the conditional distribution for mixture indicator variables, denoted $z_{iv}$, which indicates component $h$ with conditional probability

\begin{align*}
    & P(z_{iv}=h|-) = \frac{o_h}{\sum_{m=1}^H o_m}, o_m = \pi_{i,m} \left(s_{im}^2\right)^{\frac{|S_{iv}|}{2}} \exp\left[-\frac{1}{2s_{im}^2} \sum_{j \in S_{iv}} \mathcal{R}_{ij}^2(v) \right]
\end{align*}

\noindent where $S_{iv}$ denotes the set of subjects $j$ at site $i$ and voxel $v$, which may not be equal for all $v$ due to some voxels being screened out or unobserved.

Then we update the prior weights for component assignment, $\boldsymbol{\pi}_i$ using the conditional posterior

\begin{align*}
    & \boldsymbol{\pi}_i \sim \text{Dirichlet}(N_{i,1}+a_{\pi}/H, \ldots, N_{i,H}+a_{\pi}/H)
\end{align*}

\noindent where $N_{i,h}$ is the number of voxels $v$ for which $z_{iv}=h$.

Lastly, we update the noise variance components $s_{ih}^2$ using the conditional posterior

\begin{align*}
    & s_{ih}^2 \sim \text{IG}\left(a_{\epsilon}+\frac{1}{2} \sum_j |\Omega_{ij,h}|, b_{\epsilon}+\frac{1}{2} \sum_j \sum_{v \in \Omega_{ij,h}} \mathcal{R}_{ij}(v)^2 \right)
\end{align*}

\noindent where $\Omega_{ij,h}$ is the set of observed voxels for site $i$ and subject $j$ such that $z_{iv}=h$.

We set $\sigma_{\epsilon,iv}^2 = s_{ih}^2$ if $z_{iv}=h$. This leads to at most $H$ unique noise variance components for a given site $i$, allowing for model parsimony.


\subsection{ADNI dataset and preprocessing}

The Alzheimer's Disease Neuroimaging Initiative (ADNI) is a multisite consortium aimed at identifying neuroimaging-based biomarkers for Alzheimer's Disease (AD) and other neurodegenerative disorders \cite{mueller2005alzheimer}. Throughout its two-decade time span, ADNI has provided open access to large neuroimaging datasets that have been used to examine longitudinal changes in brain structure and function, assess effectiveness of treatment interventions in various populations, and predict cognitive and behavioral factors from images, among other applications \cite{weiner2010alzheimer}. For this study, we obtained $2108$ 1.5T T1-weighted magnetic resonance images (T1-w MRI) from ADNI-1, which correspond to $818$ distinct subjects with one baseline visit and between 1 and 2 follow-up visits. Images were collected over $58$ study sites, where each site had a single MRI scanner and each subject visited a single study site for all their longitudinal visits. Refer Table \ref{tab:adni_summ} for the list of sites and corresponding sample sizes). The time between baseline and the first visit was 6 months on average, while the time between baseline and the second visits was 12 months on average. In addition to neuroimaging data, the ADNI-1 dataset also consists of demographic variables, including age, gender, and ethnicity, and clinical information such as diagnosis of Alzheimer's Disease (AD), late mild cognitive impairment (LMCI), or healthy control (CN). Some limited genetic information is provided for each subject, namely the number of alleles of the apolipoprotein E gene (APOE4). Additionally, a series of cognitive tests are performed for each subject, including the Mini Mental State Examination (MMSE), Rey's Auditory Verbal Learning Test (RAVLT), and Alzheimer's Disease Assessment Scale (ADAS). Such cognitive tests have been shown to be useful predictors of AD onset and progression \cite{choe2020mmse,maheux2023forecasting,moradi2017rey}. 

All selected T1-weighted MRI scans were preprocessed in R using the ANTs toolbox \cite{tustison2021antsx}. As a first step, each scan was registered to a population template image generated from all subject images, which is recommended for AD cohorts with potentially varying brain sizes due to neurodegeneration \cite{wu2012registration}. In estimating this population template image, both the template intensity map and template tissue segmentation labels were obtained, which are provided as a part of the ANTs pipeline. Following registration, cortical thickness (CT) measures were obtained using the ANTs cortical thickness pipeline on the registered images, and subsequently data harmonization was performed on the CT measurements. 

\vspace{.25cm}

{\noindent \underline{Down-sampling images and screening voxels:}} 
Prior to cortical thickness extraction, each T1w-MRI scan was cropped to remove regions outside of the brain mask for all subjects, and images was subsequently downsampled to a size of $30 \times 30 \times 30$, or 27,000 voxels per image, using built-in ANTsR functionality (``resampleImage''). In doing so, we allow for more efficient harmonization and downstream analyses, while still preserving a relatively high resolution. Other resolutions were considered, but did not show significant benefits compared to the chosen image size, and came at the cost of substantially higher run-times. 

Since cortical thickness measures are only defined within the cerebral cortex, the outputted images were sparse, consisting of approximately 85\% zero voxels per image (ranges from 80.3\%-89.4\% across images). Additionally, even with image registration, a certain proportion of voxels in the cortical thickness maps were observed to be nonzero for some subjects and zero for others. To address this issue and allow for more stable model fitting, we screened out certain voxels from all subjects based on the proportion of subjects who had zero cortical thickness for a given value. After some preliminary testing, we chose to screen out voxels which were zero for over 75\% of images (i.e. $\geq 1,581$ images out of $2108$). Of the $30$ z-slices, $29$ were observed to have at least 5\% observed voxels remaining after screening.

\vspace{.25cm}

{\noindent \underline{Cortical Thickness Extraction:} }Cortical thickness (CT) estimates were obtained using the ANTs cortical thickness pipeline on the registered and downsampled T1-weighted MRI images. The CT extraction pipeline involves first applying N4 bias field correction to the registered images and subsequently performing state-of-the-art brain extraction within a segmentation framework \cite{tustison2021antsx}. Then, the processed brain images, estimated brain masks, and template tissue labels were used to run 6-tissue Atropos segmentation and generate tissue masks for cerebrospinal fluid (CSF), gray matter (GM), white matter (WM), deep gray matter (DGM), brain stem, and cerebellum. In this step, the tissue masks from the template image act as priors which inform the segmentation for each observation scan. Lastly, cortical thickness measurements were obtained using the DiReCT algorithm \cite{avants2014insight}, which uses tissue labels and probability maps for the GM and WM to estimate cortical thickness at each image voxel using an iterative optimization approach.

\vspace{.25cm}

{\noindent \underline{Obtaining ROI-level estimates: } We use the LONI Probabilistic Brain Atlas (LPBA) to convert voxel-level images (both unharmonized and harmonized) to ROI-level images. Particularly, LPBA contains a sub-atlas of $56$ ROIs within the gray matter cortex, which we use for cortical thickness images. A list of all LPBA GM ROIs is included in Supplementary Table \ref{tab:roi_volume}. The ROIs were used in downstream analysis for interpreting significant brain regions after the voxel-level harmonization was performed.}

\vspace{.25cm}

{\noindent \underline{Explanatory variables included in harmonization pipeline:}} The full set of variables accounted for during harmonization included demographic features (i.e. age and gender), clinical measures (i.e. diagnosis of AD, diagnosis of LMCI, and the number of APOE-4 alleles), and months since baseline visit for the longitudinal harmonization models. Cognitive scores such as Mini Mental State Exam (MMSE) score and Rey Auditory Visual Learning Test (RAVLT) score were not included in harmonization models, but rather explored in downstream analyses after harmonization was performed. This allowed us to examine if potentially relevant biological signals were adequately retained following harmonization even without explicitly accounting for those associations in the harmonization model. From a neurobiological perspective, including such cognitive scores within a regression framework that models an image outcome would violate the expected direction of causality, since the brain changes were expected to affect the cognitive scores and not the other way around.


\subsection{Details of statistical analysis with Tensor-ComBat harmonization}

To evaluate the performance of the proposed harmonization model, we examined several downstream metrics scanner effect quantification, prediction, and biological feature selection. For each aspect, we compared the performance of Tensor-ComBat, ComBat, adjusted residuals, and unharmonized. Each harmonization method consisted of a cross-sectional and longitudinal variation, and for Tensor-ComBat in particular, we compared the performance of harmonizing the full 3D images directly with harmonizing each 2D slice separately.

\subsubsection{Scanner Effect Removal}
For assessment of scanner effect removal, we first computed the scanner-level averages in the voxel-wise unharmonized and harmonized images, and use these average images to find pairwise correlations and RMSEs over all $\binom{58}{2} = 1653$ pairs of scanners. Optimal scanner effect removal should produce distributions showing higher correlations and lower RMSE across scanner pairs. In addition to pairwise assessment, we performed a series of ANOVA and Bartlett tests to determine which features in the unharmonized and harmonized images had evidence for significant differences in mean and variance, respectively. We performed these tests for aggregate summary measures derived from the unharmonized and harmonized images (i.e. mean and median) and additionally performed the test separately for each voxel in the brain mask. Under this scheme, scanner removal can be quantified by determining if the proportion of voxels showing significant differences in scanner-level mean and variance is significantly lower than the unharmonized case. In addition to performing the aforementioned tests on the imaging data, we repeated the tests on the set of tensor-valued residuals, where residuals were obtained by fitting a BTRR model on each harmonized outcome controlling for biological covariates.

We characterized the spatial location and extent of estimated scanner effects using the Tensor-ComBat MCMC samples. In particular, we computed spatial significance maps for the additive and multiplicative scanner effects using joint credible bands \cite{bonferroni1936teoria}. 
Based on the areas showing high degrees of significant scanner deviations, we then compared the distributions of scanner deviations to competing harmonization methods across the entire cortical mask and in sub-regions.

\subsubsection{Biological Analysis}
Given that harmonization is geared towards assisting in downstream biological analyses, the bulk of our evaluation criteria were rooted in biological prediction, effect estimation, and reproducibility. For prediction, we compared the out-of-sample performance for predicting biological covariates including age, disease status, and cognitive scores, using each harmonized image set as competing feature spaces. For each type of harmonized image, we ran a 5-fold cross-validation scheme predicting each of the 5 biological variables (Age, gender, AD, LMCI, APOE-4) included in the harmonization model using information from the (un)harmonized imaging features. Models were fitting using Lasso regression, where tuning parameters were determined with a separate 10-fold cross-validation using all samples. We also repeated the corresponding prediction results corresponding to Elastic Net (EN) regression in the Supplementary Materials that show similar trends as the Lasso analysis (results not presented). Additionally, we ran similar prediction schemes predicting 3 cognitive measures (MMSE, RAVLT, ADAS), which were not directly included in the harmonization model. The predictive performance of the cognitive variables were reported separately for each longitudinal visit.

We then assessed reproducibility of biological effects estimation across longitudinal visits. To evaluate this component, we stratified the harmonized images by longitudinal visit (baseline, 6-month follow-up, and 12-month follow-up) and fit a set of Bayesian tensor response regression models (BTRR) to ascertain the effects of 5 given biological covariates on cortical thickness. Models were fit on both stratified batches of data as well as the entire dataset. Then, for each effect, significance estimates were obtained using joint credible intervals \cite{bonferroni1936teoria}. This procedure allowed for built-in multiplicity correction leveraging information across voxels and MCMC samples when calculating credible intervals. We compared the spatial similarity of significance estimates between each data split and the full dataset estimate using the Dice scores, which quantify the degree of spatial overlap between two image signals \cite{dice1945measures}. 

\printbibliography
\newpage
\section{Supplementary Materials}

\begin{table}[h!]
    \centering
    \begin{tabular}{|l|c|c|c|}
    \hline
    \multirow{2}{*}{\textbf{Cortical ROI}} & \multirow{2}{*}{\textbf{\% of Volume}} & \multicolumn{2}{c|}{\textbf{\% of Scanner pairs w/ signif. deviations}} \\
    & & \textbf{Unadjusted (wilcoxon)} & \textbf{Adjusted (TC-Sl)} \\
    \hline
    Brain Stem & 4.15 & 15.9 & 4.66 \\
    Cerebellum & 11.3 & 27.5 & 35.8 \\
    \hline
    L Angular Gyrus & 1.74 & 11.4 & 2.10 \\
    R Angular Gyrus & 1.38 & 11.4 & 1.17 \\
    \hline
    L Caudate & 1.39 & 10.9 & 4.30 \\
    R Caudate & 0.795 & 9.56 & 3.90 \\
    \hline
    L Cingulate Gyrus & 1.53 & 3.33 & 2.10  \\
    R Cingulate Gyrus & 1.09 & 5.02 & 2.09  \\
    \hline
    L Cuneus & 0.866 & 9.07 & 3.64 \\
    R Cuneus & 0.681 & 14.3 & 0.614 \\
    \hline
    L Fusiform Gyrus & 1.26 & 12.6 & 9.31 \\
    R Fusiform Gyrus & 0.878 & 11.3 & 7.58 \\
    \hline
    L Gyrus Rectus & 0.703 & 17.0 & 8.51 \\
    R Gyrus Rectus & 0.401 & 6.84 & 3.40 \\
    \hline
    L Hippocampus & 0.847 & 9.20 & 6.19 \\
    R Hippocampus & 0.485 & 10.6 & 6.67 \\
    \hline
    L Inferior Frontal Gyrus & 1.98 & 16.1 & 4.07 \\
    R Inferior Frontal Gyrus & 1.80 & 12.6 & 3.73 \\
    \hline
    L Inferior Occipital Gyrus & 1.35 & 9.50 & 7.83 \\
    R Inferior Occipital Gyrus & 1.09 & 16.8 & 8.53 \\
    \hline
    L Inferior Temporal Gyrus & 1.99 & 14.4 & 10.7 \\
    R Inferior Temporal Gyrus & 1.77 & 14.8 & 11.5 \\
    \hline
    L Insular Cortex & 1.41 & 11.7 & 3.22 \\
    R Insular Cortex & 1.15 & 13.9 & 4.67 \\
    \hline
    L Lateral Orbitofrontal Gyrus & 1.10 & 9.62 & 3.27  \\
    R Lateral Orbitofrontal Gyrus & 0.773 & 11.1 & 4.17  \\
    \hline
    L Lingual Gyrus & 2.08 & 7.68 & 2.24  \\
    R Lingual Gyrus & 1.77 & 13.9 & 2.33 \\
    \hline
    L Middle Frontal Gyrus & 3.74 & 14.2 & 2.60 \\
    R Middle Frontal Gyrus & 3.56 & 11.0 & 1.73 \\
    \hline
    L Middle Occipital Gyrus & 2.30 & 8.71 & 4.24 \\
    R Middle Occipital Gyrus & 2.13 & 12.6 & 3.86 \\
    \hline
    L Middle Orbitofrontal Gyrus & 1.65 & 9.68 & 4.84  \\
    R Middle Orbitofrontal Gyrus & 1.39 & 11.9 & 5.76  \\
    \hline
    L Middle Temporal Gyrus & 2.03 & 12.0 & 6.90 \\
    R Middle Temporal Gyrus & 2.05 & 12.2 & 6.07 \\
    \hline
    L Parahippocampal Gyrus & 0.885 & 9.80 & 4.62 \\
    R Parahippocampal Gyrus & 0.773 & 5.70 & 5.70 \\
    \hline
    L Postcentral Gyrus & 1.78 & 12.3 & 1.23 \\
    R Postcentral Gyrus & 1.72 & 10.1 & 0.781 \\
    \hline
    L Precentral Gyrus & 2.80 & 13.4 & 0.841 \\
    R Precentral Gyrus & 2.64 & 11.0 & 0.436 \\
    \hline
    L Precuneus & 1.06 & 4.84 & 0.769 \\
    R Precuneus & 0.747 & 7.38 & 0.604 \\
    \hline
    L Putamen & 0.987 & 13.3 & 3.86 \\
    R Putamen & 0.752 & 11.7 & 3.36 \\
    \hline
    L Superior Frontal Gyrus & 3.56 & 11.1 & 1.07 \\
    R Superior Frontal Gyrus & 3.20 & 11.4 & 1.49 \\
    \hline
    L Superior Occipital Gyrus & 1.24 & 5.32 & 1.02 \\
    R Superior Occipital Gyrus & 1.04 & 10.2 & 1.19 \\
    \hline
    L Superior Parietal Gyrus & 2.23 & 7.02 & 0.559 \\
    R Superior Parietal Gyrus & 1.94 & 7.08 & 0.490 \\
    \hline
    L Superior Temporal Gyrus & 2.31 & 14.6 & 3.65 \\
    R Superior Temporal Gyrus & 1.93 & 10.5 & 2.56 \\
    \hline
    L Supramarginal Gyrus & 1.06 & 12.0 & 1.96 \\
    R Supramarginal Gyrus & 0.715 & 7.56 & 0.471 \\
    \hline
    \end{tabular}
    \caption{ROI Volume as Percentage of Total Volume and proportion of scanner pairs with significant additive deviations, averaged across voxels within each ROI. Two methods are used for testing for significant scanner differences, namely: an unadjusted approach using Wilcoxon log-rank test with Bonferroni correction to test for significant cortical thickness differences among scanner pairs, and an adjusted approach using a TC-3D harmonization model controlling for biological covariates and spatial dependence, where significance estimates for pairwise scanner differences are obtained using joint credible bands.}
    \label{tab:roi_volume}
\end{table}

\begin{table}[ht]
\centering
\begin{tabular}{|c|cc|cc|}
  \hline
  & \multicolumn{2}{|c|}{Full Dataset} & \multicolumn{2}{|c|}{50\% Subset} \\
Comparison & p (unadj.) & p (adj.) & p (unadj.) & p (adj.) \\ 
  \hline
TC-CS-Sl vs. TC-CS-3D & 2.55e-43 & 8.17e-42 & 1.5e-06 & 5.08e-05 \\ 
  TC-CS-Sl vs. C-CS & 3.03e-20 & 8.47e-19 & 0.00161 & 0.0386 \\ 
  TC-CS-Sl vs. AR-CS & 2.31e-31 & 6.93e-30 & 5.38e-05 & 0.00145 \\ 
  TC-CS-Sl vs. TC-L-Sl & 3.41e-49 & 1.13e-47 & 0.000285 & 0.00712 \\ 
  TC-CS-Sl vs. TC-L-3D & 1.05e-52 & 3.67e-51 & 2.26e-05 & 0.000679 \\ 
  TC-CS-Sl vs. C-L & 2.35e-62 & 8.46e-61 & 1.4e-08 & 5.04e-07 \\ 
  TC-CS-Sl vs. AR-L & 3.54e-38 & 1.1e-36 & 9.15e-06 & 0.000284 \\ 
  TC-CS-Sl vs. U & 3.21e-52 & 1.09e-50 & 4.68e-06 & 0.00015 \\ 
  TC-CS-3D vs. C-CS & 1.41e-09 & 3.25e-08 & 0.861 & 0.861 \\ 
  TC-CS-3D vs. AR-CS & 8.13e-05 & 0.00114 & 0.111 & 0.861 \\ 
  TC-CS-3D vs. TC-L-Sl & 0.948 & 0.948 & 0.558 & 0.861 \\ 
  TC-CS-3D vs. TC-L-3D & 0.118 & 0.941 & 0.203 & 0.861 \\ 
  TC-CS-3D vs. C-L & 1.88e-06 & 3.19e-05 & 2.16e-06 & 7.12e-05 \\ 
  TC-CS-3D vs. AR-L & 0.436 & 0.948 & 0.00935 & 0.168 \\ 
  TC-CS-3D vs. U & 0.522 & 0.948 & 0.0041 & 0.0879 \\ 
  C-CS vs. AR-CS & 0.0115 & 0.126 & 0.274 & 0.861 \\ 
  C-CS vs. TC-L-Sl & 1.05e-10 & 2.52e-09 & 0.58 & 0.861 \\ 
  C-CS vs. TC-L-3D & 1.48e-14 & 3.86e-13 & 0.448 & 0.861 \\ 
  C-CS vs. C-L & 2.11e-24 & 6.13e-23 & 4.5e-05 & 0.00126 \\ 
  C-CS vs. AR-L & 2.41e-07 & 4.58e-06 & 0.0453 & 0.725 \\ 
  C-CS vs. U & 1.18e-12 & 2.96e-11 & 0.0243 & 0.414 \\ 
  AR-CS vs. TC-L-Sl & 2.42e-05 & 0.000363 & 0.0671 & 0.861 \\ 
  AR-CS vs. TC-L-3D & 1.41e-08 & 3.11e-07 & 0.648 & 0.861 \\ 
  AR-CS vs. C-L & 1.31e-17 & 3.53e-16 & 0.000281 & 0.00712 \\ 
  AR-CS vs. AR-L & 0.00244 & 0.0293 & 0.287 & 0.861 \\ 
  AR-CS vs. U & 7.71e-07 & 1.39e-05 & 0.173 & 0.861 \\ 
  TC-L-Sl vs. TC-L-3D & 0.077 & 0.693 & 0.119 & 0.861 \\ 
  TC-L-Sl vs. C-L & 2.13e-07 & 4.25e-06 & 1.24e-06 & 4.33e-05 \\ 
  TC-L-Sl vs. AR-L & 0.438 & 0.948 & 0.00591 & 0.118 \\ 
  TC-L-Sl vs. U & 0.439 & 0.948 & 0.00262 & 0.0603 \\ 
  TC-L-3D vs. C-L & 0.000562 & 0.0073 & 3.98e-05 & 0.00115 \\ 
  TC-L-3D vs. AR-L & 0.0203 & 0.203 & 0.118 & 0.861 \\ 
  TC-L-3D vs. U & 0.3 & 0.948 & 0.0621 & 0.861 \\ 
  C-L vs. AR-L & 7.73e-08 & 1.62e-06 & 0.00418 & 0.0879 \\ 
  C-L vs. U & 5.78e-06 & 9.24e-05 & 0.00816 & 0.155 \\ 
  AR-L vs. U & 0.145 & 0.948 & 0.76 & 0.861 \\ 
   \hline
\end{tabular}
\caption{P-values and Adjusted P-values for \textbf{Age} Prediction (CV-5), including samples from all 3 visits. Adjusted p-values obtained with Benjamini-Hochberg correction over the 36 pairs of methods being compared.}
\label{tab:agePred_s}
\end{table}

\begin{table}[ht]
\centering
\begin{tabular}{|c|cc|cc|cc|}
  \hline
  & \multicolumn{2}{|c|}{Baseline} & \multicolumn{2}{|c|}{6-Month Follow-up} & \multicolumn{2}{|c|}{12-Month Follow-up} \\
Comparison & p (unadj.) & p (adj.) & p (unadj.) & p (adj.) & p (unadj.) & p (adj.) \\ 
  \hline
TC-CS-Sl vs. TC-CS-3D & 2.14e-14 & 6.84e-13 & 3.56e-17 & 1.1e-15 & 3.73e-16 & 1.19e-14 \\ 
  TC-CS-Sl vs. C-CS & 2.06e-13 & 5.97e-12 & 1.06e-07 & 2.74e-06 & 7.08e-06 & 0.000191 \\ 
  TC-CS-Sl vs. AR-CS & 1.86e-13 & 5.59e-12 & 2.67e-12 & 8.02e-11 & 4.72e-11 & 1.42e-09 \\ 
  TC-CS-Sl vs. TC-L-Sl & 4.1e-18 & 1.35e-16 & 1.58e-18 & 5.06e-17 & 5.54e-18 & 1.94e-16 \\ 
  TC-CS-Sl vs. TC-L-3D & 5.48e-19 & 1.92e-17 & 8.16e-22 & 2.77e-20 & 1.37e-17 & 4.52e-16 \\ 
  TC-CS-Sl vs. C-L & 1.63e-26 & 5.87e-25 & 1.85e-23 & 6.65e-22 & 3.14e-19 & 1.13e-17 \\ 
  TC-CS-Sl vs. AR-L & 1.77e-13 & 5.48e-12 & 1.42e-18 & 4.7e-17 & 2.41e-11 & 7.48e-10 \\ 
  TC-CS-Sl vs. U & 1.02e-18 & 3.47e-17 & 1.3e-22 & 4.56e-21 & 6.77e-18 & 2.3e-16 \\ 
  TC-CS-3D vs. C-CS & 0.0762 & 0.87 & 4.82e-06 & 0.000106 & 1.53e-05 & 0.000351 \\ 
  TC-CS-3D vs. AR-CS & 0.139 & 0.87 & 0.00568 & 0.0964 & 0.0075 & 0.13 \\ 
  TC-CS-3D vs. TC-L-Sl & 0.766 & 0.87 & 0.852 & 0.96 & 0.838 & 0.992 \\ 
  TC-CS-3D vs. TC-L-3D & 0.234 & 0.87 & 0.463 & 0.96 & 0.403 & 0.992 \\ 
  TC-CS-3D vs. C-L & 3.63e-05 & 0.000871 & 0.0173 & 0.225 & 0.0424 & 0.551 \\ 
  TC-CS-3D vs. AR-L & 0.87 & 0.87 & 0.815 & 0.96 & 0.363 & 0.992 \\ 
  TC-CS-3D vs. U & 0.175 & 0.87 & 0.933 & 0.96 & 0.846 & 0.992 \\ 
  C-CS vs. AR-CS & 0.765 & 0.87 & 0.0338 & 0.406 & 0.0401 & 0.551 \\ 
  C-CS vs. TC-L-Sl & 0.0162 & 0.275 & 3.04e-06 & 7.3e-05 & 7.43e-06 & 0.000192 \\ 
  C-CS vs. TC-L-3D & 0.00115 & 0.0242 & 1.87e-08 & 5.23e-07 & 7.2e-07 & 2.02e-05 \\ 
  C-CS vs. C-L & 2.73e-10 & 7.64e-09 & 2.4e-11 & 6.96e-10 & 7.73e-09 & 2.24e-07 \\ 
  C-CS vs. AR-L & 0.123 & 0.87 & 3.35e-06 & 7.71e-05 & 0.00207 & 0.0434 \\ 
  C-CS vs. U & 0.000788 & 0.0181 & 1.05e-07 & 2.74e-06 & 7.7e-06 & 0.000192 \\ 
  AR-CS vs. TC-L-Sl & 0.0415 & 0.664 & 0.00603 & 0.0964 & 0.00635 & 0.122 \\ 
  AR-CS vs. TC-L-3D & 0.00376 & 0.0676 & 0.000175 & 0.00368 & 0.000587 & 0.0129 \\ 
  AR-CS vs. C-L & 2.85e-09 & 7.7e-08 & 3.43e-07 & 8.58e-06 & 8.77e-06 & 0.000211 \\ 
  AR-CS vs. AR-L & 0.207 & 0.87 & 0.00667 & 0.0991 & 0.148 & 0.992 \\ 
  AR-CS vs. U & 0.00255 & 0.051 & 0.00127 & 0.0254 & 0.00641 & 0.122 \\ 
  TC-L-Sl vs. TC-L-3D & 0.299 & 0.87 & 0.325 & 0.96 & 0.268 & 0.992 \\ 
  TC-L-Sl vs. C-L & 1.07e-05 & 0.000277 & 0.00708 & 0.0991 & 0.018 & 0.282 \\ 
  TC-L-Sl vs. AR-L & 0.637 & 0.87 & 0.96 & 0.96 & 0.426 & 0.992 \\ 
  TC-L-Sl vs. U & 0.22 & 0.87 & 0.894 & 0.96 & 0.992 & 0.992 \\ 
  TC-L-3D vs. C-L & 0.00104 & 0.0229 & 0.0623 & 0.686 & 0.222 & 0.992 \\ 
  TC-L-3D vs. AR-L & 0.182 & 0.87 & 0.298 & 0.96 & 0.102 & 0.992 \\ 
  TC-L-3D vs. U & 0.833 & 0.87 & 0.323 & 0.96 & 0.274 & 0.992 \\ 
  C-L vs. AR-L & 2.85e-05 & 0.000713 & 0.00595 & 0.0964 & 0.00762 & 0.13 \\ 
  C-L vs. U & 0.00277 & 0.0527 & 0.00396 & 0.0753 & 0.0188 & 0.282 \\ 
  AR-L vs. U & 0.136 & 0.87 & 0.849 & 0.96 & 0.423 & 0.992 \\ 
   \hline
\end{tabular}
\caption{P-values and Adjusted P-values for \textbf{Age} Prediction (CV-5), stratified by longitudinal visit. Adjusted p-values obtained with Benjamini-Hochberg correction over the 36 pairs of methods being compared.}
\end{table}

\begin{table}[ht]
\centering
\begin{tabular}{|c|cc|cc|}
  \hline
  & \multicolumn{2}{|c|}{Full Dataset} & \multicolumn{2}{|c|}{50\% Subset} \\
Comparison & p (unadj.) & p (adj.) & p (unadj.) & p (adj.) \\ 
  \hline
TC-CS-Sl vs. TC-CS-3D & 0.0133 & 0.143 & 0.00261 & 0.0679 \\ 
  TC-CS-Sl vs. C-CS & 0.976 & 0.976 & 0.996 & 0.996 \\ 
  TC-CS-Sl vs. AR-CS & 0.681 & 0.976 & 0.406 & 0.996 \\ 
  TC-CS-Sl vs. TC-L-Sl & 8.35e-11 & 2.51e-09 & 0.677 & 0.996 \\ 
  TC-CS-Sl vs. TC-L-3D & 1.89e-10 & 5.48e-09 & 0.214 & 0.996 \\ 
  TC-CS-Sl vs. C-L & 4.55e-12 & 1.46e-10 & 0.000678 & 0.021 \\ 
  TC-CS-Sl vs. AR-L & 0.00231 & 0.0347 & 0.149 & 0.996 \\ 
  TC-CS-Sl vs. U & 1.01e-05 & 0.000212 & 0.0572 & 0.914 \\ 
  TC-CS-3D vs. C-CS & 0.38 & 0.976 & 0.00562 & 0.124 \\ 
  TC-CS-3D vs. AR-CS & 0.718 & 0.976 & 0.000806 & 0.0242 \\ 
  TC-CS-3D vs. TC-L-Sl & 1.15e-11 & 3.58e-10 & 4.24e-06 & 0.000153 \\ 
  TC-CS-3D vs. TC-L-3D & 3.53e-12 & 1.17e-10 & 0.00135 & 0.0364 \\ 
  TC-CS-3D vs. C-L & 2.17e-12 & 7.36e-11 & 7.83e-06 & 0.000274 \\ 
  TC-CS-3D vs. AR-L & 0.000208 & 0.00374 & 0.000627 & 0.02 \\ 
  TC-CS-3D vs. U & 7.41e-07 & 1.67e-05 & 0.000141 & 0.00467 \\ 
  C-CS vs. AR-CS & 0.76 & 0.976 & 0.429 & 0.996 \\ 
  C-CS vs. TC-L-Sl & 1.69e-07 & 4.23e-06 & 0.704 & 0.996 \\ 
  C-CS vs. TC-L-3D & 5.6e-05 & 0.00112 & 0.264 & 0.996 \\ 
  C-CS vs. C-L & 9.3e-13 & 3.35e-11 & 0.000891 & 0.025 \\ 
  C-CS vs. AR-L & 0.0143 & 0.143 & 0.165 & 0.996 \\ 
  C-CS vs. U & 0.000333 & 0.00566 & 0.068 & 0.996 \\ 
  AR-CS vs. TC-L-Sl & 6.37e-07 & 1.53e-05 & 0.499 & 0.996 \\ 
  AR-CS vs. TC-L-3D & 0.000117 & 0.00222 & 0.0458 & 0.869 \\ 
  AR-CS vs. C-L & 1.16e-12 & 4.07e-11 & 0.00427 & 0.102 \\ 
  AR-CS vs. AR-L & 0.0113 & 0.135 & 0.483 & 0.996 \\ 
  AR-CS vs. U & 0.000377 & 0.00603 & 0.255 & 0.996 \\ 
  TC-L-Sl vs. TC-L-3D & 0.00657 & 0.0854 & 0.00423 & 0.102 \\ 
  TC-L-Sl vs. C-L & 7.6e-07 & 1.67e-05 & 0.000855 & 0.0248 \\ 
  TC-L-Sl vs. AR-L & 0.0028 & 0.0392 & 0.165 & 0.996 \\ 
  TC-L-Sl vs. U & 0.0497 & 0.447 & 0.0554 & 0.914 \\ 
  TC-L-3D vs. C-L & 5.13e-09 & 1.39e-07 & 0.000118 & 0.00401 \\ 
  TC-L-3D vs. AR-L & 0.215 & 0.976 & 0.0176 & 0.37 \\ 
  TC-L-3D vs. U & 0.872 & 0.976 & 0.00463 & 0.107 \\ 
  C-L vs. AR-L & 9.41e-10 & 2.64e-08 & 0.0277 & 0.555 \\ 
  C-L vs. U & 1.32e-08 & 3.44e-07 & 0.0513 & 0.914 \\ 
  AR-L vs. U & 0.249 & 0.976 & 0.698 & 0.996 \\ 
   \hline
\end{tabular}
\caption{P-values and Adjusted P-values for \textbf{Gender} Prediction (CV-5). Adjusted p-values obtained with Benjamini-Hochberg correction over the 36 pairs of methods being compared.}
\label{tab:genderPred_s}
\end{table}

\begin{table}[ht]
\centering
\begin{tabular}{|c|cc|cc|}
  \hline
  & \multicolumn{2}{|c|}{Full Dataset} & \multicolumn{2}{|c|}{50\% Subset} \\
Comparison & p (unadj.) & p (adj.) & p (unadj.) & p (adj.) \\ 
  \hline
TC-CS-Sl vs. TC-CS-3D & 2.62e-06 & 4.71e-05 & 0.249 & 0.991 \\ 
  TC-CS-Sl vs. C-CS & 3.22e-07 & 6.75e-06 & 0.0544 & 0.991 \\ 
  TC-CS-Sl vs. AR-CS & 1.8e-07 & 4.31e-06 & 0.0225 & 0.653 \\ 
  TC-CS-Sl vs. TC-L-Sl & 3.02e-21 & 1.09e-19 & 0.0021 & 0.0713 \\ 
  TC-CS-Sl vs. TC-L-3D & 1.81e-13 & 5.97e-12 & 0.00481 & 0.159 \\ 
  TC-CS-Sl vs. C-L & 5.16e-14 & 1.75e-12 & 0.000249 & 0.00898 \\ 
  TC-CS-Sl vs. AR-L & 6.74e-13 & 2.16e-11 & 0.0128 & 0.383 \\ 
  TC-CS-Sl vs. U & 1.12e-09 & 3.15e-08 & 0.00914 & 0.293 \\ 
  TC-CS-3D vs. C-CS & 0.254 & 0.746 & 0.245 & 0.991 \\ 
  TC-CS-3D vs. AR-CS & 0.0401 & 0.281 & 0.112 & 0.991 \\ 
  TC-CS-3D vs. TC-L-Sl & 5.06e-14 & 1.75e-12 & 0.00985 & 0.305 \\ 
  TC-CS-3D vs. TC-L-3D & 6.87e-09 & 1.85e-07 & 0.0244 & 0.684 \\ 
  TC-CS-3D vs. C-L & 6.85e-11 & 2.05e-09 & 0.000896 & 0.0314 \\ 
  TC-CS-3D vs. AR-L & 6.43e-05 & 0.00103 & 0.0648 & 0.991 \\ 
  TC-CS-3D vs. U & 0.000412 & 0.00536 & 0.0463 & 0.991 \\ 
  C-CS vs. AR-CS & 0.304 & 0.746 & 0.728 & 0.991 \\ 
  C-CS vs. TC-L-Sl & 6.5e-11 & 2.02e-09 & 0.241 & 0.991 \\ 
  C-CS vs. TC-L-3D & 2.67e-07 & 5.87e-06 & 0.431 & 0.991 \\ 
  C-CS vs. C-L & 7.19e-10 & 2.08e-08 & 0.0473 & 0.991 \\ 
  C-CS vs. AR-L & 0.00591 & 0.065 & 0.558 & 0.991 \\ 
  C-CS vs. U & 0.00989 & 0.0989 & 0.465 & 0.991 \\ 
  AR-CS vs. TC-L-Sl & 1.6e-07 & 3.99e-06 & 0.399 & 0.991 \\ 
  AR-CS vs. TC-L-3D & 2.2e-05 & 0.000374 & 0.671 & 0.991 \\ 
  AR-CS vs. C-L & 2.08e-08 & 5.4e-07 & 0.0864 & 0.991 \\ 
  AR-CS vs. AR-L & 0.153 & 0.746 & 0.806 & 0.991 \\ 
  AR-CS vs. U & 0.132 & 0.746 & 0.691 & 0.991 \\ 
  TC-L-Sl vs. TC-L-3D & 0.704 & 0.746 & 0.624 & 0.991 \\ 
  TC-L-Sl vs. C-L & 0.0122 & 0.107 & 0.311 & 0.991 \\ 
  TC-L-Sl vs. AR-L & 4.89e-07 & 9.78e-06 & 0.555 & 0.991 \\ 
  TC-L-Sl vs. U & 9.17e-05 & 0.00138 & 0.67 & 0.991 \\ 
  TC-L-3D vs. C-L & 0.0134 & 0.107 & 0.134 & 0.991 \\ 
  TC-L-3D vs. AR-L & 0.000294 & 0.00412 & 0.875 & 0.991 \\ 
  TC-L-3D vs. U & 0.00251 & 0.0302 & 0.991 & 0.991 \\ 
  C-L vs. AR-L & 2.52e-07 & 5.81e-06 & 0.136 & 0.991 \\ 
  C-L vs. U & 1.94e-06 & 3.69e-05 & 0.181 & 0.991 \\ 
  AR-L vs. U & 0.746 & 0.746 & 0.878 & 0.991 \\ 
   \hline
\end{tabular}
\caption{P-values and Adjusted P-values for \textbf{AD} Prediction (CV-5). Adjusted p-values obtained with Benjamini-Hochberg correction over the 36 pairs of methods being compared.}
\label{tab:adPred_s}
\end{table}

\begin{table}[ht]
\centering
\begin{tabular}{|c|cc|cc|}
  \hline
  & \multicolumn{2}{|c|}{Full Dataset} & \multicolumn{2}{|c|}{50\% Subset} \\
Comparison & p (unadj.) & p (adj.) & p (unadj.) & p (adj.) \\ 
  \hline
TC-CS-Sl vs. TC-CS-3D & 0.0591 & 0.355 & 0.893 & 0.893 \\ 
  TC-CS-Sl vs. C-CS & 0.0218 & 0.196 & 0.491 & 0.893 \\ 
  TC-CS-Sl vs. AR-CS & 5.19e-05 & 0.000727 & 0.168 & 0.893 \\ 
  TC-CS-Sl vs. TC-L-Sl & 8.01e-13 & 2.32e-11 & 0.196 & 0.893 \\ 
  TC-CS-Sl vs. TC-L-3D & 1.98e-22 & 7.14e-21 & 0.00237 & 0.0735 \\ 
  TC-CS-Sl vs. C-L & 1.71e-20 & 5.8e-19 & 0.000255 & 0.00891 \\ 
  TC-CS-Sl vs. AR-L & 1.28e-08 & 2.56e-07 & 0.0428 & 0.893 \\ 
  TC-CS-Sl vs. U & 3.34e-10 & 7.69e-09 & 0.0242 & 0.63 \\ 
  TC-CS-3D vs. C-CS & 0.313 & 0.627 & 0.387 & 0.893 \\ 
  TC-CS-3D vs. AR-CS & 0.0016 & 0.0192 & 0.108 & 0.893 \\ 
  TC-CS-3D vs. TC-L-Sl & 8.34e-11 & 2.08e-09 & 0.122 & 0.893 \\ 
  TC-CS-3D vs. TC-L-3D & 4.95e-21 & 1.73e-19 & 0.000974 & 0.0321 \\ 
  TC-CS-3D vs. C-L & 1.23e-19 & 4.07e-18 & 0.000105 & 0.00379 \\ 
  TC-CS-3D vs. AR-L & 8.3e-07 & 1.33e-05 & 0.0258 & 0.645 \\ 
  TC-CS-3D vs. U & 2.93e-08 & 5.28e-07 & 0.0139 & 0.374 \\ 
  C-CS vs. AR-CS & 0.0353 & 0.283 & 0.486 & 0.893 \\ 
  C-CS vs. TC-L-Sl & 3.6e-07 & 6.12e-06 & 0.581 & 0.893 \\ 
  C-CS vs. TC-L-3D & 4.25e-16 & 1.32e-14 & 0.00985 & 0.286 \\ 
  C-CS vs. C-L & 8.15e-18 & 2.61e-16 & 0.00097 & 0.0321 \\ 
  C-CS vs. AR-L & 0.00019 & 0.00248 & 0.14 & 0.893 \\ 
  C-CS vs. U & 2.19e-05 & 0.000329 & 0.0842 & 0.893 \\ 
  AR-CS vs. TC-L-Sl & 0.00437 & 0.048 & 0.827 & 0.893 \\ 
  AR-CS vs. TC-L-3D & 9.84e-11 & 2.36e-09 & 0.0303 & 0.728 \\ 
  AR-CS vs. C-L & 3.51e-13 & 1.05e-11 & 0.00271 & 0.0814 \\ 
  AR-CS vs. AR-L & 0.131 & 0.572 & 0.347 & 0.893 \\ 
  AR-CS vs. U & 0.0479 & 0.335 & 0.223 & 0.893 \\ 
  TC-L-Sl vs. TC-L-3D & 2.51e-08 & 4.77e-07 & 0.0132 & 0.368 \\ 
  TC-L-Sl vs. C-L & 6.68e-11 & 1.74e-09 & 0.00114 & 0.0363 \\ 
  TC-L-Sl vs. AR-L & 0.143 & 0.572 & 0.24 & 0.893 \\ 
  TC-L-Sl vs. U & 0.304 & 0.627 & 0.143 & 0.893 \\ 
  TC-L-3D vs. C-L & 0.0119 & 0.119 & 0.229 & 0.893 \\ 
  TC-L-3D vs. AR-L & 1.51e-09 & 3.33e-08 & 0.276 & 0.893 \\ 
  TC-L-3D vs. U & 2.07e-09 & 4.35e-08 & 0.409 & 0.893 \\ 
  C-L vs. AR-L & 4.53e-12 & 1.27e-10 & 0.0404 & 0.893 \\ 
  C-L vs. U & 6.68e-12 & 1.8e-10 & 0.067 & 0.893 \\ 
  AR-L vs. U & 0.633 & 0.633 & 0.797 & 0.893 \\ 
   \hline
\end{tabular}
\caption{P-values and Adjusted P-values for \textbf{LMCI} Prediction (CV-5). Adjusted p-values obtained with Benjamini-Hochberg correction over the 36 pairs of methods being compared.}
\label{tab:lmciPred_s}
\end{table}

\begin{table}[ht]
\centering
\begin{tabular}{|c|cc|cc|}
  \hline
  & \multicolumn{2}{|c|}{Full Dataset} & \multicolumn{2}{|c|}{50\% Subset} \\
Comparison & p (unadj.) & p (adj.) & p (unadj.) & p (adj.) \\ 
  \hline
TC-CS-Sl vs. TC-CS-3D & 0.179 & 0.806 & 0.23 & 0.918 \\ 
  TC-CS-Sl vs. C-CS & 0.208 & 0.806 & 0.341 & 0.918 \\ 
  TC-CS-Sl vs. AR-CS & 2.38e-08 & 5.25e-07 & 0.0938 & 0.918 \\ 
  TC-CS-Sl vs. TC-L-Sl & 1.84e-11 & 4.79e-10 & 0.219 & 0.918 \\ 
  TC-CS-Sl vs. TC-L-3D & 1.9e-13 & 5.52e-12 & 0.869 & 0.918 \\ 
  TC-CS-Sl vs. C-L & 1.64e-18 & 5.57e-17 & 1.69e-05 & 0.000576 \\ 
  TC-CS-Sl vs. AR-L & 2.06e-09 & 4.94e-08 & 0.0275 & 0.577 \\ 
  TC-CS-Sl vs. U & 3.37e-13 & 9.43e-12 & 0.0156 & 0.348 \\ 
  TC-CS-3D vs. C-CS & 0.806 & 0.806 & 0.042 & 0.799 \\ 
  TC-CS-3D vs. AR-CS & 2.26e-05 & 0.000361 & 0.00802 & 0.208 \\ 
  TC-CS-3D vs. TC-L-Sl & 2.06e-08 & 4.74e-07 & 0.0158 & 0.348 \\ 
  TC-CS-3D vs. TC-L-3D & 9.4e-08 & 1.88e-06 & 0.169 & 0.918 \\ 
  TC-CS-3D vs. C-L & 8.15e-20 & 2.93e-18 & 1.01e-06 & 3.65e-05 \\ 
  TC-CS-3D vs. AR-L & 5.41e-08 & 1.14e-06 & 0.00212 & 0.0614 \\ 
  TC-CS-3D vs. U & 2.98e-10 & 7.44e-09 & 0.000937 & 0.029 \\ 
  C-CS vs. AR-CS & 0.000953 & 0.0133 & 0.451 & 0.918 \\ 
  C-CS vs. TC-L-Sl & 5.58e-06 & 9.49e-05 & 0.882 & 0.918 \\ 
  C-CS vs. TC-L-3D & 5.51e-05 & 0.000827 & 0.187 & 0.918 \\ 
  C-CS vs. C-L & 1.95e-19 & 6.84e-18 & 0.000262 & 0.00837 \\ 
  C-CS vs. AR-L & 1.35e-06 & 2.43e-05 & 0.177 & 0.918 \\ 
  C-CS vs. U & 2.19e-07 & 4.17e-06 & 0.128 & 0.918 \\ 
  AR-CS vs. TC-L-Sl & 0.0534 & 0.48 & 0.481 & 0.918 \\ 
  AR-CS vs. TC-L-3D & 0.385 & 0.806 & 0.0341 & 0.683 \\ 
  AR-CS vs. C-L & 1.51e-16 & 4.98e-15 & 0.00175 & 0.0525 \\ 
  AR-CS vs. AR-L & 0.00642 & 0.0771 & 0.526 & 0.918 \\ 
  AR-CS vs. U & 0.00277 & 0.0361 & 0.439 & 0.918 \\ 
  TC-L-Sl vs. TC-L-3D & 0.163 & 0.806 & 0.066 & 0.918 \\ 
  TC-L-Sl vs. C-L & 6.36e-15 & 1.97e-13 & 7.02e-05 & 0.00232 \\ 
  TC-L-Sl vs. AR-L & 0.205 & 0.806 & 0.171 & 0.918 \\ 
  TC-L-Sl vs. U & 0.248 & 0.806 & 0.114 & 0.918 \\ 
  TC-L-3D vs. C-L & 2.22e-15 & 7.1e-14 & 1.89e-06 & 6.61e-05 \\ 
  TC-L-3D vs. AR-L & 0.0185 & 0.185 & 0.00871 & 0.218 \\ 
  TC-L-3D vs. U & 0.00807 & 0.0887 & 0.00334 & 0.0934 \\ 
  C-L vs. AR-L & 1.66e-12 & 4.47e-11 & 0.00994 & 0.239 \\ 
  C-L vs. U & 6.57e-14 & 1.97e-12 & 0.00789 & 0.208 \\ 
  AR-L vs. U & 0.725 & 0.806 & 0.918 & 0.918 \\ 
   \hline
\end{tabular}
\caption{P-values and Adjusted P-values for \textbf{APOE-4} Prediction (CV-5). Adjusted p-values obtained with Benjamini-Hochberg correction over the 36 pairs of methods being compared.}
\label{tab:apoePred_s}
\end{table}

\begin{table}[ht]
\centering
\begin{tabular}{|c|cc|cc|cc|}
  \hline
  & \multicolumn{2}{|c|}{Baseline} & \multicolumn{2}{|c|}{6-Month Follow-up} & \multicolumn{2}{|c|}{12-Month Follow-up} \\
Comparison & p (unadj.) & p (adj.) & p (unadj.) & p (adj.) & p (unadj.) & p (adj.) \\ 
  \hline
TC-CS-Sl vs. TC-CS-3D & 6.55e-05 & 0.0021 & 0.167 & 0.973 & 0.0448 & 0.961 \\ 
  TC-CS-Sl vs. C-CS & 0.000137 & 0.00412 & 0.212 & 0.973 & 0.139 & 0.961 \\ 
  TC-CS-Sl vs. AR-CS & 7.94e-07 & 2.78e-05 & 0.107 & 0.973 & 0.0628 & 0.961 \\ 
  TC-CS-Sl vs. TC-L-Sl & 9.45e-05 & 0.00293 & 0.0215 & 0.731 & 0.178 & 0.961 \\ 
  TC-CS-Sl vs. TC-L-3D & 0.00251 & 0.0729 & 0.234 & 0.973 & 0.492 & 0.961 \\ 
  TC-CS-Sl vs. C-L & 1.83e-08 & 6.59e-07 & 0.0198 & 0.692 & 0.025 & 0.875 \\ 
  TC-CS-Sl vs. AR-L & 3.94e-06 & 0.000134 & 0.0804 & 0.973 & 0.0381 & 0.961 \\ 
  TC-CS-Sl vs. U & 2.04e-05 & 0.000673 & 0.0133 & 0.479 & 0.0174 & 0.627 \\ 
  TC-CS-3D vs. C-CS & 0.95 & 0.993 & 0.973 & 0.973 & 0.528 & 0.961 \\ 
  TC-CS-3D vs. AR-CS & 0.0761 & 0.993 & 0.664 & 0.973 & 0.72 & 0.961 \\ 
  TC-CS-3D vs. TC-L-Sl & 0.993 & 0.993 & 0.258 & 0.973 & 0.611 & 0.961 \\ 
  TC-CS-3D vs. TC-L-3D & 0.834 & 0.993 & 0.86 & 0.973 & 0.252 & 0.961 \\ 
  TC-CS-3D vs. C-L & 0.0109 & 0.304 & 0.198 & 0.973 & 0.625 & 0.961 \\ 
  TC-CS-3D vs. AR-L & 0.0955 & 0.993 & 0.561 & 0.973 & 0.891 & 0.961 \\ 
  TC-CS-3D vs. U & 0.132 & 0.993 & 0.164 & 0.973 & 0.769 & 0.961 \\ 
  C-CS vs. AR-CS & 0.106 & 0.993 & 0.727 & 0.973 & 0.745 & 0.961 \\ 
  C-CS vs. TC-L-Sl & 0.945 & 0.993 & 0.361 & 0.973 & 0.961 & 0.961 \\ 
  C-CS vs. TC-L-3D & 0.803 & 0.993 & 0.855 & 0.973 & 0.533 & 0.961 \\ 
  C-CS vs. C-L & 0.0202 & 0.524 & 0.272 & 0.973 & 0.292 & 0.961 \\ 
  C-CS vs. AR-L & 0.124 & 0.993 & 0.634 & 0.973 & 0.453 & 0.961 \\ 
  C-CS vs. U & 0.162 & 0.993 & 0.25 & 0.973 & 0.332 & 0.961 \\ 
  AR-CS vs. TC-L-Sl & 0.0812 & 0.993 & 0.592 & 0.973 & 0.82 & 0.961 \\ 
  AR-CS vs. TC-L-3D & 0.108 & 0.993 & 0.567 & 0.973 & 0.356 & 0.961 \\ 
  AR-CS vs. C-L & 0.54 & 0.993 & 0.442 & 0.973 & 0.405 & 0.961 \\ 
  AR-CS vs. AR-L & 0.956 & 0.993 & 0.9 & 0.973 & 0.621 & 0.961 \\ 
  AR-CS vs. U & 0.971 & 0.993 & 0.426 & 0.973 & 0.48 & 0.961 \\ 
  TC-L-Sl vs. TC-L-3D & 0.842 & 0.993 & 0.211 & 0.973 & 0.546 & 0.961 \\ 
  TC-L-Sl vs. C-L & 0.0126 & 0.341 & 0.737 & 0.973 & 0.362 & 0.961 \\ 
  TC-L-Sl vs. AR-L & 0.0999 & 0.993 & 0.688 & 0.973 & 0.535 & 0.961 \\ 
  TC-L-Sl vs. U & 0.136 & 0.993 & 0.741 & 0.973 & 0.427 & 0.961 \\ 
  TC-L-3D vs. C-L & 0.0296 & 0.741 & 0.163 & 0.973 & 0.14 & 0.961 \\ 
  TC-L-3D vs. AR-L & 0.119 & 0.993 & 0.474 & 0.973 & 0.216 & 0.961 \\ 
  TC-L-3D vs. U & 0.147 & 0.993 & 0.134 & 0.973 & 0.148 & 0.961 \\ 
  C-L vs. AR-L & 0.617 & 0.993 & 0.515 & 0.973 & 0.723 & 0.961 \\ 
  C-L vs. U & 0.638 & 0.993 & 0.969 & 0.973 & 0.802 & 0.961 \\ 
  AR-L vs. U & 0.989 & 0.993 & 0.503 & 0.973 & 0.889 & 0.961 \\ 
   \hline
\end{tabular}
\caption{P-values and Adjusted P-values for \textbf{MMSE} Prediction (CV-5) over three longitudinal visits. Adjusted p-values obtained with Benjamini-Hochberg correction over the 36 pairs of methods being compared.}
\label{tab:mmsePred_s}
\end{table}

\begin{table}[ht]
\centering
\begin{tabular}{|c|cc|cc|cc|}
  \hline
  & \multicolumn{2}{|c|}{Baseline} & \multicolumn{2}{|c|}{6-Month Follow-up} & \multicolumn{2}{|c|}{12-Month Follow-up} \\
Comparison & p (unadj.) & p (adj.) & p (unadj.) & p (adj.) & p (unadj.) & p (adj.) \\ 
  \hline
TC-CS-Sl vs. TC-CS-3D & 0.121 & 0.977 & 0.242 & 0.963 & 0.38 & 0.943 \\ 
  TC-CS-Sl vs. C-CS & 0.000949 & 0.0294 & 0.000672 & 0.0235 & 0.00133 & 0.0439 \\ 
  TC-CS-Sl vs. AR-CS & 0.000564 & 0.0181 & 0.000838 & 0.0285 & 0.0014 & 0.0448 \\ 
  TC-CS-Sl vs. TC-L-Sl & 0.00979 & 0.284 & 6.25e-05 & 0.00225 & 0.00479 & 0.139 \\ 
  TC-CS-Sl vs. TC-L-3D & 6.58e-05 & 0.0023 & 0.00151 & 0.0467 & 0.00269 & 0.0806 \\ 
  TC-CS-Sl vs. C-L & 2.36e-05 & 0.000848 & 0.00118 & 0.0376 & 5.77e-05 & 0.00208 \\ 
  TC-CS-Sl vs. AR-L & 0.000323 & 0.0107 & 0.000949 & 0.0313 & 0.0017 & 0.0527 \\ 
  TC-CS-Sl vs. U & 0.000223 & 0.00759 & 0.00172 & 0.0516 & 0.000861 & 0.0301 \\ 
  TC-CS-3D vs. C-CS & 0.0406 & 0.975 & 0.0409 & 0.963 & 0.015 & 0.39 \\ 
  TC-CS-3D vs. AR-CS & 0.0285 & 0.712 & 0.0369 & 0.963 & 0.0148 & 0.39 \\ 
  TC-CS-3D vs. TC-L-Sl & 0.262 & 0.977 & 0.00899 & 0.261 & 0.0552 & 0.943 \\ 
  TC-CS-3D vs. TC-L-3D & 0.0111 & 0.31 & 0.0579 & 0.963 & 0.0293 & 0.703 \\ 
  TC-CS-3D vs. C-L & 0.00263 & 0.079 & 0.0403 & 0.963 & 0.001 & 0.0341 \\ 
  TC-CS-3D vs. AR-L & 0.0283 & 0.712 & 0.0299 & 0.836 & 0.0201 & 0.502 \\ 
  TC-CS-3D vs. U & 0.0233 & 0.629 & 0.0707 & 0.963 & 0.00938 & 0.263 \\ 
  C-CS vs. AR-CS & 0.906 & 0.977 & 0.847 & 0.963 & 0.943 & 0.943 \\ 
  C-CS vs. TC-L-Sl & 0.296 & 0.977 & 0.519 & 0.963 & 0.376 & 0.943 \\ 
  C-CS vs. TC-L-3D & 0.962 & 0.977 & 0.951 & 0.963 & 0.709 & 0.943 \\ 
  C-CS vs. C-L & 0.4 & 0.977 & 0.816 & 0.963 & 0.367 & 0.943 \\ 
  C-CS vs. AR-L & 0.908 & 0.977 & 0.652 & 0.963 & 0.824 & 0.943 \\ 
  C-CS vs. U & 0.926 & 0.977 & 0.809 & 0.963 & 0.772 & 0.943 \\ 
  AR-CS vs. TC-L-Sl & 0.237 & 0.977 & 0.697 & 0.963 & 0.35 & 0.943 \\ 
  AR-CS vs. TC-L-3D & 0.927 & 0.977 & 0.811 & 0.963 & 0.663 & 0.943 \\ 
  AR-CS vs. C-L & 0.467 & 0.977 & 0.963 & 0.963 & 0.422 & 0.943 \\ 
  AR-CS vs. AR-L & 0.804 & 0.977 & 0.794 & 0.963 & 0.772 & 0.943 \\ 
  AR-CS vs. U & 0.819 & 0.977 & 0.681 & 0.963 & 0.831 & 0.943 \\ 
  TC-L-Sl vs. TC-L-3D & 0.184 & 0.977 & 0.508 & 0.963 & 0.613 & 0.943 \\ 
  TC-L-Sl vs. C-L & 0.0462 & 0.977 & 0.751 & 0.963 & 0.0548 & 0.943 \\ 
  TC-L-Sl vs. AR-L & 0.29 & 0.977 & 0.949 & 0.963 & 0.496 & 0.943 \\ 
  TC-L-Sl vs. U & 0.267 & 0.977 & 0.381 & 0.963 & 0.248 & 0.943 \\ 
  TC-L-3D vs. C-L & 0.333 & 0.977 & 0.783 & 0.963 & 0.188 & 0.943 \\ 
  TC-L-3D vs. AR-L & 0.841 & 0.977 & 0.629 & 0.963 & 0.874 & 0.943 \\ 
  TC-L-3D vs. U & 0.86 & 0.977 & 0.868 & 0.963 & 0.513 & 0.943 \\ 
  C-L vs. AR-L & 0.285 & 0.977 & 0.834 & 0.963 & 0.243 & 0.943 \\ 
  C-L vs. U & 0.286 & 0.977 & 0.66 & 0.963 & 0.571 & 0.943 \\ 
  AR-L vs. U & 0.977 & 0.977 & 0.519 & 0.963 & 0.609 & 0.943 \\ 
   \hline
\end{tabular}
\caption{P-values and Adjusted P-values for \textbf{ADAS} Prediction (CV-5) over three longitudinal visits. Adjusted p-values obtained with Benjamini-Hochberg correction over the 36 pairs of methods being compared.}
\label{tab:adasPred_s}
\end{table}

\begin{table}[ht]
\centering
\begin{tabular}{|c|cc|cc|cc|}
  \hline
  & \multicolumn{2}{|c|}{Baseline} & \multicolumn{2}{|c|}{6-Month Follow-up} & \multicolumn{2}{|c|}{12-Month Follow-up} \\
Comparison & p (unadj.) & p (adj.) & p (unadj.) & p (adj.) & p (unadj.) & p (adj.) \\ 
  \hline
TC-CS-Sl vs. TC-CS-3D & 0.305 & 0.957 & 0.708 & 0.996 & 0.612 & 0.971 \\ 
  TC-CS-Sl vs. C-CS & 0.00397 & 0.139 & 0.0249 & 0.748 & 0.151 & 0.971 \\ 
  TC-CS-Sl vs. AR-CS & 0.0373 & 0.957 & 0.0229 & 0.711 & 0.147 & 0.971 \\ 
  TC-CS-Sl vs. TC-L-Sl & 0.237 & 0.957 & 0.0016 & 0.0577 & 0.0468 & 0.971 \\ 
  TC-CS-Sl vs. TC-L-3D & 0.00919 & 0.312 & 0.0123 & 0.392 & 0.589 & 0.971 \\ 
  TC-CS-Sl vs. C-L & 0.00256 & 0.092 & 0.00618 & 0.216 & 0.00259 & 0.0934 \\ 
  TC-CS-Sl vs. AR-L & 0.0248 & 0.818 & 0.0287 & 0.834 & 0.122 & 0.971 \\ 
  TC-CS-Sl vs. U & 0.0277 & 0.888 & 0.00768 & 0.261 & 0.121 & 0.971 \\ 
  TC-CS-3D vs. C-CS & 0.109 & 0.957 & 0.0991 & 0.996 & 0.338 & 0.971 \\ 
  TC-CS-3D vs. AR-CS & 0.301 & 0.957 & 0.095 & 0.996 & 0.322 & 0.971 \\ 
  TC-CS-3D vs. TC-L-Sl & 0.863 & 0.957 & 0.0112 & 0.371 & 0.116 & 0.971 \\ 
  TC-CS-3D vs. TC-L-3D & 0.135 & 0.957 & 0.0634 & 0.996 & 0.955 & 0.971 \\ 
  TC-CS-3D vs. C-L & 0.047 & 0.957 & 0.0362 & 0.996 & 0.00778 & 0.272 \\ 
  TC-CS-3D vs. AR-L & 0.226 & 0.957 & 0.101 & 0.996 & 0.273 & 0.971 \\ 
  TC-CS-3D vs. U & 0.257 & 0.957 & 0.0404 & 0.996 & 0.268 & 0.971 \\ 
  C-CS vs. AR-CS & 0.679 & 0.957 & 0.995 & 0.996 & 0.931 & 0.971 \\ 
  C-CS vs. TC-L-Sl & 0.169 & 0.957 & 0.278 & 0.996 & 0.466 & 0.971 \\ 
  C-CS vs. TC-L-3D & 0.957 & 0.957 & 0.857 & 0.996 & 0.396 & 0.971 \\ 
  C-CS vs. C-L & 0.472 & 0.957 & 0.62 & 0.996 & 0.0586 & 0.971 \\ 
  C-CS vs. AR-L & 0.851 & 0.957 & 0.944 & 0.996 & 0.837 & 0.971 \\ 
  C-CS vs. U & 0.748 & 0.957 & 0.632 & 0.996 & 0.812 & 0.971 \\ 
  AR-CS vs. TC-L-Sl & 0.396 & 0.957 & 0.275 & 0.996 & 0.539 & 0.971 \\ 
  AR-CS vs. TC-L-3D & 0.674 & 0.957 & 0.861 & 0.996 & 0.375 & 0.971 \\ 
  AR-CS vs. C-L & 0.326 & 0.957 & 0.621 & 0.996 & 0.0893 & 0.971 \\ 
  AR-CS vs. AR-L & 0.848 & 0.957 & 0.948 & 0.996 & 0.909 & 0.971 \\ 
  AR-CS vs. U & 0.932 & 0.957 & 0.633 & 0.996 & 0.883 & 0.971 \\ 
  TC-L-Sl vs. TC-L-3D & 0.196 & 0.957 & 0.339 & 0.996 & 0.149 & 0.971 \\ 
  TC-L-Sl vs. C-L & 0.0728 & 0.957 & 0.53 & 0.996 & 0.301 & 0.971 \\ 
  TC-L-Sl vs. AR-L & 0.305 & 0.957 & 0.334 & 0.996 & 0.618 & 0.971 \\ 
  TC-L-Sl vs. U & 0.346 & 0.957 & 0.536 & 0.996 & 0.651 & 0.971 \\ 
  TC-L-3D vs. C-L & 0.542 & 0.957 & 0.739 & 0.996 & 0.0137 & 0.464 \\ 
  TC-L-3D vs. AR-L & 0.829 & 0.957 & 0.923 & 0.996 & 0.323 & 0.971 \\ 
  TC-L-3D vs. U & 0.736 & 0.957 & 0.749 & 0.996 & 0.316 & 0.971 \\ 
  C-L vs. AR-L & 0.433 & 0.957 & 0.691 & 0.996 & 0.116 & 0.971 \\ 
  C-L vs. U & 0.362 & 0.957 & 0.996 & 0.996 & 0.136 & 0.971 \\ 
  AR-L vs. U & 0.913 & 0.957 & 0.701 & 0.996 & 0.971 & 0.971 \\ 
   \hline
\end{tabular}
\caption{P-values and Adjusted P-values for \textbf{RAVLT} Prediction (CV-5) over three longitudinal visits. Adjusted p-values obtained with Benjamini-Hochberg correction over the 36 pairs of methods being compared.}
\label{tab:ravltPred_s}
\end{table}

\begin{table}[ht]
\centering
\begin{tabular}{|c|cc|cc|}
  \hline
  & \multicolumn{2}{|c|}{Age} & \multicolumn{2}{|c|}{Gender} \\
Comparison & p (unadj.) & p (adj.) & p (unadj.) & p (adj.) \\ 
  \hline
TC-CS-Sl vs. TC-CS-3D & 0.000352 & 0.00633 & 0.00149 & 0.0182 \\ 
  TC-CS-Sl vs. C-CS & 0.131 & 0.658 & 0.000515 & 0.00823 \\ 
  TC-CS-Sl vs. TC-L-Sl & 0.876 & 0.876 & 0.137 & 0.582 \\ 
  TC-CS-Sl vs. TC-L-3D & 0.00265 & 0.0345 & 0.000661 & 0.00992 \\ 
  TC-CS-Sl vs. C-L & 0.154 & 0.658 & 0.438 & 0.582 \\ 
  TC-CS-Sl vs. U & 0.274 & 0.823 & 0.0299 & 0.209 \\ 
  TC-CS-3D vs. C-CS & 0.000293 & 0.00558 & 2.82e-09 & 5.93e-08 \\ 
  TC-CS-3D vs. TC-L-Sl & 8.37e-05 & 0.00167 & 0.0272 & 0.209 \\ 
  TC-CS-3D vs. TC-L-3D & 0.0846 & 0.658 & 0.181 & 0.582 \\ 
  TC-CS-3D vs. C-L & 0.000728 & 0.0122 & 0.00198 & 0.0198 \\ 
  TC-CS-3D vs. U & 4.01e-06 & 8.42e-05 & 1.49e-07 & 2.98e-06 \\ 
  C-CS vs. TC-L-Sl & 0.148 & 0.658 & 0.582 & 0.582 \\ 
  C-CS vs. TC-L-3D & 0.000883 & 0.0132 & 1.92e-06 & 3.65e-05 \\ 
  C-CS vs. C-L & 0.0135 & 0.162 & 4.5e-06 & 8.1e-05 \\ 
  C-CS vs. U & 0.0228 & 0.228 & 0.00192 & 0.0198 \\ 
  TC-L-Sl vs. TC-L-3D & 0.000762 & 0.0122 & 0.0138 & 0.125 \\ 
  TC-L-Sl vs. C-L & 0.089 & 0.658 & 0.097 & 0.582 \\ 
  TC-L-Sl vs. U & 0.164 & 0.658 & 0.309 & 0.582 \\ 
  TC-L-3D vs. C-L & 0.0204 & 0.224 & 0.00152 & 0.0182 \\ 
  TC-L-3D vs. U & 0.00101 & 0.0141 & 1.88e-05 & 0.00032 \\ 
  C-L vs. U & 0.521 & 0.876 & 0.00074 & 0.0104 \\ 
   \hline
\end{tabular}
\caption{Comparison of Dice scores for reproducibility of significance estimates for \textbf{age} and \textbf{gender} on harmonized cortical thickness outcomes. Ten replicated estimates were obtained using random 50\% subsets of the data, and compared to the estimates obtained using the full dataset.}
\label{tab:agegender_rep}
\end{table}

\begin{table}[ht]
\centering
\begin{tabular}{|c|cc|cc|}
  \hline
  & \multicolumn{2}{|c|}{AD} & \multicolumn{2}{|c|}{LMCI} \\
Comparison & p (unadj.) & p (adj.) & p (unadj.) & p (adj.) \\ 
  \hline
TC-CS-Sl vs. TC-CS-3D & 0.000122 & 0.00196 & 0.000446 & 0.00893 \\ 
  TC-CS-Sl vs. C-CS & 0.143 & 0.669 & 0.589 & 0.882 \\ 
  TC-CS-Sl vs. TC-L-Sl & 0.000221 & 0.00309 & 0.0755 & 0.755 \\ 
  TC-CS-Sl vs. TC-L-3D & 0.358 & 0.903 & 0.0736 & 0.755 \\ 
  TC-CS-Sl vs. C-L & 0.167 & 0.669 & 0.535 & 0.882 \\ 
  TC-CS-Sl vs. U & 0.0207 & 0.145 & 0.585 & 0.882 \\ 
  TC-CS-3D vs. C-CS & 4.47e-06 & 8.94e-05 & 0.000144 & 0.00303 \\ 
  TC-CS-3D vs. TC-L-Sl & 4.27e-07 & 8.96e-06 & 0.000921 & 0.0175 \\ 
  TC-CS-3D vs. TC-L-3D & 0.000155 & 0.00232 & 0.0191 & 0.286 \\ 
  TC-CS-3D vs. C-L & 0.00234 & 0.0272 & 0.00116 & 0.0208 \\ 
  TC-CS-3D vs. U & 0.00291 & 0.0291 & 0.0165 & 0.264 \\ 
  C-CS vs. TC-L-Sl & 0.00125 & 0.0162 & 0.113 & 0.882 \\ 
  C-CS vs. TC-L-3D & 0.903 & 0.903 & 0.0222 & 0.311 \\ 
  C-CS vs. C-L & 0.00568 & 0.0511 & 0.231 & 0.882 \\ 
  C-CS vs. U & 6.36e-05 & 0.00108 & 0.369 & 0.882 \\ 
  TC-L-Sl vs. TC-L-3D & 0.00248 & 0.0272 & 0.0138 & 0.235 \\ 
  TC-L-Sl vs. C-L & 3.41e-05 & 0.000647 & 0.0441 & 0.573 \\ 
  TC-L-Sl vs. U & 3.64e-05 & 0.000656 & 0.054 & 0.648 \\ 
  TC-L-3D vs. C-L & 0.0544 & 0.326 & 0.19 & 0.882 \\ 
  TC-L-3D vs. U & 0.0142 & 0.114 & 0.458 & 0.882 \\ 
  C-L vs. U & 0.458 & 0.903 & 0.882 & 0.882 \\ 
   \hline
\end{tabular}
\caption{Comparison of Dice scores for reproducibility of significance estimates for \textbf{AD} and \textbf{LMCI} on harmonized cortical thickness outcomes. Ten replicated estimates were obtained using random 50\% subsets of the data, and compared to the estimates obtained using the full dataset.}
\label{tab:adlmci_rep}
\end{table}

\end{document}